\documentclass[12pt]{article}
\usepackage{setspace}

\usepackage{url}
\usepackage[utf8]{inputenc}
\usepackage{amsmath}
\usepackage{latexsym}
\usepackage{graphicx,color}
\usepackage[usenames,dvipsnames,svgnames,table]{xcolor}
\usepackage{bm}
\usepackage{bbm,amssymb,epsfig}
\RequirePackage{amsthm,amsmath,amssymb,amsfonts,graphicx,subfigure,bm}
\usepackage{float}
\usepackage{soul}
\usepackage{enumitem}
\usepackage{makecell}

\usepackage{natbib}
\usepackage{xcolor}
\usepackage{multirow}
\usepackage{ulem}

\usepackage{lipsum}
\usepackage{enumitem}
\usepackage{ulem}
\usepackage{colortbl}
\usepackage{multirow}
\usepackage[margin = 1in]{geometry}
\usepackage{graphicx,psfrag,epsfig,epstopdf-base}
\usepackage{comment}
\usepackage{wrapfig}
\usepackage{subfigure}
\usepackage{tikz}
\usepackage{geometry}
\usepackage{pdflscape}
\usepackage{epsfig}
\usepackage{graphicx}
\usepackage{amsfonts}
\usepackage{hyperref}
\usepackage{latexsym}
\usepackage{epstopdf}
\usepackage{mathtools}
\usepackage{amsmath,amssymb,amsthm}
\usepackage{bm}
\usepackage{bbm}
\usepackage{booktabs}
\usepackage{parskip}
\usepackage[ruled,vlined]{algorithm2e}
\usepackage{algorithmic}
\usepackage{cleveref, autonum}

\newcommand{\GP}{\mathrm{GP}}

\newcommand{\iid}{\stackrel{\rm iid}{\sim}}
\newcommand{\ind}{\textbf{1}}

\newcommand{\RR}{\mathbb{R}}
\newcommand{\PP}{\mathbb{P}}
\newcommand{\EE}{\mathbb{E}}
\newcommand{\s}{\sum_{i=1}^{\infty}}
\newcommand{\la}{\lambda}

\RequirePackage[OT1]{fontenc}

\usepackage{mathrsfs}

\newcommand{\bbH}{\mathbb{H}}
\newcommand{\mX}{\mathcal{X}}

\newcommand\by{\bm{y}}

\newcommand{\bI}{\bm{I}}

\newcommand{\Ltwo}{L^2_{p_X}(\mX)}

\newcommand{\erf}{\,{\rm Erf}}

\newcommand{\E}{\mathrm{E}}
\newcommand{\Var}{\mathrm{Var}}

\newtheorem{thm}{Theorem}

\newtheorem{lem}{Lemma}
\newtheorem{prop}{Proposition}

\theoremstyle{definition} 
\theoremstyle{definition} 
\newtheorem{remark}{Remark}

\graphicspath{{figures/}}
\begin{document}

\title{Semiparametric Bayesian inference for local extrema of functions in the presence of noise}

\author{Meng Li, Zejian Liu, Cheng-Han Yu and Marina Vannucci} 
\date{}
\vspace{-1in}
\maketitle

\begin{abstract}
There is a wide range of applications where the local extrema of a function are the key quantity of interest. However, there is surprisingly little work on methods to infer local extrema with uncertainty quantification in the presence of noise. By viewing the function as an infinite-dimensional nuisance parameter, a semiparametric formulation of this problem poses daunting challenges, both methodologically and theoretically, as (i) the number of local extrema may be unknown, and (ii) the induced shape constraints associated with local extrema are highly irregular. In this article, we build upon a derivative-constrained Gaussian process prior recently proposed by \cite{yu2023bayesian} to derive what we call an encompassing approach that indexes possibly multiple local extrema by a single parameter. We provide closed-form characterization of the posterior distribution and study its large sample behavior under this unconventional encompassing regime. We show that the posterior measure converges to a mixture of Gaussians with the number of components matching the underlying truth, leading to posterior exploration that accounts for multi-modality. Point and interval estimates of local extrema with frequentist properties are also provided. The encompassing approach leads to a remarkably simple, fast semiparametric approach for inference on local extrema. We illustrate the method through simulations and a real data application to event-related potential analysis.
\end{abstract}
\vspace{-0.2in}
\noindent
{\bf Keywords:} Local extrema, Gaussian process, semiparametric, shape-constrained regression, Bernstein-von Mises theorem

\newpage

\allowdisplaybreaks[4]

\section{Introduction}

Finding localized features of a smooth function, including local maxima and local minima, plays a pervasive role in statistics, with wide-ranging scientific applications such as in biology~\citep{raghuraman2001replication}, microscopy~\citep{egner2007fluorescence,geisler2007resolution}, and psychology~\citep{Luck2005}. Moreover, localized features provide additional characterizations of the shape of a function that are useful for visualization and interpretation, and lead to insights into optimization, particularly when operated on approximations of the function.

There is a rich literature on shape-constrained regression, where the overwhelming emphasis has been on incorporating restrictions, including monotonicity, convexity, modality, log-concavity, and piecewise constants, within nonparametric modeling of the underlying surface; see, for example,  \cite{Ramsay1998, Holmes2003, Neelon2004, Meyer2008, Shively2009, Shively2011, Abraham2015,wheeler2017bayesian,dasgupta2021modality}. In this article, we contribute to this growing literature by focusing on a distinct perspective, namely, the inference on local extrema that form the key characterization of the shape constraint, while the underlying regression function is less of interest and can be viewed as a nuisance parameter. 

There is surprisingly little work on the inference of local extrema with uncertainty quantification in the presence of noise. Notable exceptions include two-step approaches in the spirit of ``smooth first, then estimation", where one first employs nonparametric smoothing techniques, then estimates the local extrema of the smoothed estimate. Along this line, \cite{Song2006} used kernel smoothing followed by hypothesis testing to find locations at which the regression function has zero derivatives at a given statistical significance level. Since the test is performed on all locations, a multiple testing issue emerges, even when there are a limited number of local extrema. \cite{Schwartzman2011} and \cite{Cheng2017} studied false discovery rate control and power consistency for local maxima under a unimodal true peak assumption. However, uncertainty quantification of the detected local maxima is not reported. Alternatively to the two-step approach, \cite{davies2001local} proposed to use the taut string method for piecewise monotone functions, and \cite{Kovac2007} extended the approach to smooth functions for finding point estimates of local extrema. 

In this article, we consider a semiparametric Bayesian method for local extrema in situations where the number of local extrema may be unknown and the associated shape constraints are highly local. These pose daunting challenges to uncertainty quantification in the presence of noise, particularly when there is more than one local extremum point. Here, we build upon a derivative-constrained Gaussian process prior, recently proposed by \cite{yu2023bayesian} for the location of \textit{stationary} points in event-related potentials (ERP), to derive  what we call an \textit{encompassing} approach that indexes possibly multiple local extrema by a single parameter. 
We provide a rigorous theoretical investigation of this unconventional approach, which ensures a proper interpretation of the derived uncertainty quantification in the context of local extrema detection.
The encompassing approach is remarkably simple, as it transforms a varying-dimensional model into one dimension, and thus is particularly well suited to address the multiplicity challenge posed by local extrema. We note that we use the Bayes machinery to derive a posterior distribution but employs frequentist properties to characterize its large sample behavior and justify the obtained point and interval estimates.

In our theoretical investigation, we characterize the posterior distribution of the encompassing approach and show an intrinsic connection to unconstrained nonparametric regression, enabling fast implementation without complicated sampling. We show that the posterior measure converges to a mixture of Gaussians with the number of components matching the underlying truth. This interesting phenomenon not only provides theoretical guarantees for the inference on local extrema that accounts for multi-modality of the posterior distribution, but also extends the Bernstein-von Mises (BvM) theorem beyond the traditional semiparametric Bayesian literature to the encompassing paradigm. Classic semiparametric Bayesian BvMs typically assume separable priors on the function and finite-dimensional parameter with fixed dimension, or rely on the parameter of interest being a bounded functional of the regression function~\citep{castillo2012semiparametric,castillo2015bernstein}; we instead study the limiting posterior distribution under irregular scenarios when the local extrema have unknown dimension and are embedded in the regression function, hence not separable, and the derivative at any fixed point, when viewed as a functional of the regression function, is not bounded. This large sample characterization of the posterior distribution leads to consistent estimators of the number and location of local extrema. We additionally provide interval estimation for local extrema with frequentist coverage.

\textbf{Organization.} Section~\ref{sec:methods} introduces the model, shape-constrained priors, and a closed-form characterization of the posterior distribution of local extrema. In Section~\ref{sec:theory} we provide non-asymptotic bounds for a range of nonparametric quantities related to the posterior distribution, and establish a local asymptotic normality property and multi-modal limiting distribution under the encompassing regime. Consistent point estimators and interval estimators with frequentist coverage are also provided. In Section~\ref{sec:simulation} we carry out simulation studies, and in Section~\ref{sec:application} we illustrate the proposed method in event-related potential analysis. Section~\ref{sec:conclusion} concludes the paper. All proofs, additional technical results, and additional numerical experiments can be found in the Supplementary Materials.

\section{Methods} \label{sec:methods} 

\subsection{Shape-constrained regression} \label{sec:model} 

Suppose we observe independent and identically distributed samples $X = \{X_1, \ldots, X_n\} \in \mX^n$ and $\bm{y} = \{y_1, \ldots, y_n\} \in \mathbb{R}^n$ from a distribution $\PP_0$ on $\mX \times \RR$ with $n$ being the sample size and $\mX \subset \mathbb{R}$ the sample space for the covariate that is compact. Throughout the paper we focus on one-dimensional sample space for concreteness and ease of notation, and consider $\mX = [0, 1]$ without loss of generality. We briefly comment on extensions to the multi-dimensional space in the Discussion section.

We assume a regression model for the input data of the type
\begin{eqnarray}
y_i = f(X_i) + \epsilon _i, ~~~i = 1, \ldots, n, 
\label{eq:data_generating}
\end{eqnarray}
with $f: \mathcal{X} \rightarrow \mathbb{R}$ and $X_i \iid \PP_X$, where the measures $\PP_X$ admit a density $p_X$ with respect to the Lebesgue measure $\mu$ on $\mX$, and with random noise $\epsilon_i \iid N(0, \sigma ^ 2)$. 

We make the following assumptions on the true regression function $f_0$.

\textbf{Assumption A1.} $f_0\in C^2(\mX)$.

\textbf{Assumption A2.} $f_0$ has exactly $M$ local extrema for some finite $M \geq 1$ at $0 < t_1 < \cdots < t_M < 1$. 

\textbf{Assumption A3.} $f''_0(t_m) \neq 0$ for $m = 1, \ldots, M$.

Assumption A1 trivially implies that $f_0$ is bounded since it is continuous on a closed interval. Assumption A2 means that $f_0$ possesses exactly $M$ local extrema, with $M \geq 1$ finite but unknown, and local extrema do not occur at the boundary of $\mX$. Assumptions A1 and A2 lead to a necessary condition for $t_m$ to be a local extremum: $f'_0(t_m) = 0$ for $m = 1, \ldots, M.$ Assumption A3 regularizes the curvature of $f_0$ at each local extremum. Assumptions A2 and A3 indicate that we focus on local extrema that can be identified based on the second derivative test. Unlike some existing work such as \cite{davies2001local}, we do not assume that all stationary points (zeros of $f_0'$) are local extrema, and our assumptions do not regularize stationary points that are not local extrema.

Our goal is to make inference on $\{t_1, \ldots, t_M\}$ when $M$ is unknown, with uncertainty quantification. As such, we next proceed to constrained priors on $f$ accounting for local extrema.

\subsection{Shape-constrained prior of $f$ on local extrema}
The underlying function $f$ is unknown, and its local extrema are encoded in the function derivatives. We follow \cite{yu2023bayesian} and adopt a constrained Gaussian process prior under derivative constraints.

A widely used prior for $f$ is a Gaussian process (GP) with mean 0 and a covariance kernel that determines its key properties. Starting with a covariance kernel $k(\cdot,\cdot) = \sigma^2 (n \lambda)^{-1} K(\cdot,\cdot)$, where $K:\mX\times\mX\rightarrow\RR$ is a continuous, symmetric and positive definite bivariate function, we encode the derivative constraint by conditioning this GP prior on $f'(t) = 0$ for an unknown scalar parameter $t$. Assuming differentiability of $K(\cdot, \cdot)$, let $K_{jl}(x,x')={\partial ^{j+l}K(x,x')}/{\partial x^{j}\partial x'^{l}}$ for any $j, l \geq 0$. Then by direct calculation, the conditional GP is also a GP with mean 0 and covariance kernel $k_t(x, x') = \sigma^2 (n \lambda)^{-1} \{K(x, x') - K_{01}(x, t) K_{11}^{-1}(t, t) K_{10}(t, x')\},$ provided that $K_{11}(t, t) > 0$. The tuning parameter $\lambda$ possibly depends on the sample size $n$. Later we will make all assumptions on $K(\cdot, \cdot)$ clear. 

Under the constrained prior $\GP(0, k_t)$, the sample path $f(\cdot)$ satisfies $\E(f'(t)) = 0$ and 
\[
\Var(f'(t)) = \left.\frac{\partial k_t^2(x, x')}{\partial x \partial x'} \right|_{(x, x') = (t, t)} = \sigma^2 (n \lambda)^{-1} \{K_{11}(t, t) - K_{11}(t, t) K_{11}^{-1}(t, t) K_{11}(t, t)\} = 0. 
\]
Hence, it holds that $f'(t) = 0$ almost surely. Note that here we employ the differentiability of sample paths of Gaussian processes with continuously differentiable covariance kernels and the covariance function of $f'$ that is induced by differentiating $k_t$~\citep[e.g., see][Proposition I.3]{Ghosal+van:17}.

We conclude the specification of all priors by placing a prior $\pi(t)$ on $t$, which is supported on $\mX$. Thus, the marginal prior distribution of $f$ is a mixture of constrained GPs if we integrate out $t$ with respect to its prior distribution. 

Like \cite{yu2023bayesian}, we use a univariate $t$ to index all possible local extrema. This \textit{encompassing} strategy eliminates the need to specify the number of local extrema, which is particularly useful when $M$ is unknown and possibly greater than one. In addition, it enables unified inference on \textit{all} local extrema through the Bayes machinery. Although practically appealing, this unconventional encompassing regime in a semiparametric setting is not well understood in the literature. \cite{yu2023bayesian}, in particular, employ Monte Carlo EM to conduct an empirical exploration of the posterior of $t$. A specific focus of this article is a rigorous characterization of the induced posterior distribution, both at finite sample size and asymptotically, which is critical to interpret and substantiate such a strategy, while providing insights into how to carry out posterior summary.

\subsection{Closed-form posterior distribution of $t$} \label{sec:closed.form} 

Integrating out $f$ in Model~\eqref{eq:data_generating} with respect to its prior $ \GP(0, k_t)$ gives the marginal distribution 
$\bm{y} | X,t \sim N(0, \Sigma_t)$ with  
\begin{align}
\Sigma_t  &= \sigma^2 (n \lambda)^{-1}\left\{K(X, X) -  K_{01}(X, t)K_{11}^{-1}(t, t)K_{10}(t, X) \right\} + \sigma^2 \bI_n\\ 
&= \{\sigma^2 (n \lambda)^{-1}K(X, X) + \sigma^2 \bI_n\} - \sigma^2 (n \lambda)^{-1}K_{01}(X, t)K_{11}^{-1}(t, t)K_{10}(t, X); \label{eq:marginal.likelihood} 
\end{align}
here $K_{01}(X, t) = (K_{01}(X_1, t), \ldots, K_{01}(X_n, t))^T$ is a length $n$ column vector, $K_{10}(t, X) = (K_{10}(t, X_1), \ldots, K_{10}(t, X_n)) = [K_{01}(X, t)]^T$ a length $n$ row vector, $K(X, X) = (K(X_i, X_j))_{i, j =1}^n$ an $n$ by $n$ matrix, and $\bI_n$ the $n$ by $n$ identity matrix. Thus, the marginal likelihood of $t$, denoted by $\ell(t) = p(\by \mid X, t)$, is the density function of $N(0, \Sigma_t)$ evaluated at $\by$. 

An intriguing observation is that the first term in Equation~\eqref{eq:marginal.likelihood} $\sigma^2 (n \lambda)^{-1}K(X, X) + \sigma^2 \bI_n$ does not depend on $t$ and coincides with the covariance matrix under the GP($0, k$) prior. This enables a reformulation of $\ell(t)$ to relate the posterior distribution of $t$ to unconstrained nonparametric regression. Before formally presenting this connection in Proposition~\ref{lem:closed-form}, we first review standard GP priors without constraints to introduce notation. 

Suppose that one uses the unconstrained GP prior $f \sim \GP(0, \sigma^2 (n \lambda)^{-1} K)$ as the prior on $f$ without shape constraints. In this article, we may omit explicit mention of the dependence on $\lambda$ in most cases, like $\phi_{11, m}$ and $\phi_{11}(\cdot),$ except for a few instances such as $f_{\lambda}$ and $t_{\lambda, m}$, which we will introduce later. By conjugacy, the posterior distribution of $f$ is also a GP: $
f | X,\by \sim \GP(\widehat{\mu}_{f}(\cdot), \widehat{\Sigma}_{f}(\cdot,\cdot)),$
where $\widehat{\mu}_{f}(x) = K(x, X) [K(X, X) + n \lambda \bI_n]^{-1} \by$ and 
$\widehat{\Sigma}_{f}(x, x') =  \sigma^2 (n \lambda)^{-1} \left\{K(x, x') - K(x, X)[K(X, X) + n \lambda \bI_n]^{-1} K(X, x')\right\}.$ Moreover, the derivative $f' |X, \by $ is also a GP with mean $\widehat{\mu}_{f'}(\cdot)$ and covariance $\widehat{\Sigma}_{f'}(\cdot, \cdot)$: 
\newcommand{\hatmu}{\widehat{\mu}}
\begin{equation}
\label{eq:mu.prime}
\widehat{\mu}_{f'}(x) = \frac{d \widehat{\mu}_f(x)}{dx} = K_{10}(x, X) [K(X, X) + n \lambda \bI_n]^{-1} \by, 
\end{equation}
and 
$
\widehat{\Sigma}_{f'}(x, x')  = \frac{\partial^2 \widehat{\Sigma}_{f}(x, x')}{\partial x \partial x'}  = \sigma^2 (n \lambda)^{-1} \left\{K_{11}(x, x') - K_{10}(x, X)[K(X, X) + n \lambda \bI_n]^{-1} K_{01}(X, x')\right\}. 
$
In particular, the marginal posterior variance of the derivative process is 
\begin{equation} \label{eq:post.var.derivative} 
\widehat{\sigma}^2_{f'}(x)=\widehat{\Sigma}_{f'}(x, x) = \sigma^2 (n \lambda)^{-1} \left\{K_{11}(x, x) - K_{10}(x, X)[K(X, X) + n \lambda \bI_n]^{-1} K_{01}(X, x)\right\}.
\end{equation}

We are now in a position to reformulate the posterior $\pi_n(t \mid X,\by)$ as follows.  
\begin{prop}\label{lem:closed-form} 
Suppose $K \in C^2(\mX, \mX)$ and $\widehat{\sigma}^2_{f'}(x) > 0$ for any $x \in \mX$. Then it holds that 
\begin{equation}\label{eq:likelihood}
\ell(t) = C \frac{1}{\sqrt{\widehat{\sigma}^2_{f'}(t)/K_{11}(t, t)}} \exp\left \{  -\frac{\widehat{\mu}_{f'}^2(t)}{2\widehat{\sigma}^2_{f'}(t)} \right \},
\end{equation}
for some constant $C$ that does not depend on $t$, where $\widehat{\mu}_{f'}^2(\cdot)$ and $\widehat{\sigma}^2_{f'}(\cdot)$ are defined in Equations~\eqref{eq:mu.prime} and~\eqref{eq:post.var.derivative}, respectively. Consequently, the posterior distribution of $t$ under the prior $\pi(t)$ satisfies
\begin{equation} \label{eq:post.closed.form} 
\pi_n(t \mid X, \by) \propto \frac{1}{\sqrt{\widehat{\sigma}^2_{f'}(t)/K_{11}(t, t)}} \exp\left( -\frac{\widehat{\mu}_{f'}^2(t)}{2\widehat{\sigma}^2_{f'}(t)} \right) \cdot \pi(t).
\end{equation}
\end{prop}

The normalizing constant in $\pi_n(t \mid X, \by)$ can be calculated using routine one-dimensional numerical integration methods, such as the midpoint or trapezoidal rule. The closed-form type of formulation for the posterior $\pi_n(t \mid X,\by)$ in  Proposition~\ref{lem:closed-form} is useful on several fronts. Computationally, a close inspection of~\eqref{eq:post.closed.form} suggests that evaluating $\pi_n(t \mid X,\by)$ in various $t$ only requires inverting an $n$ by $n$ matrix $K(X, X) + n \lambda \bI_n$ once, dramatically reducing the computation in a naive implementation that directly inverts a varying covariance matrix induced by $k_t$ at each $t$. Theoretically, Proposition~\ref{lem:closed-form} turns inference on $t$ into key quantities related to posterior inference of $f'$ with the unconstrained $\GP(0, k)$ prior, namely $\widehat{\mu}_{f'}^2(\cdot)$ and $\widehat{\sigma}^2_{f'}(\cdot)$. We next build on this connection to analyze large sample behavior of $\pi_n(t \mid X, \by)$.

\section{Theoretical results} \label{sec:theory} 

In this section, we provide theoretical evidence of a multi-modal posterior distribution of $t$. Standard Bernstein-von Mises (BvM) theorems state that, under certain conditions, the posterior distribution is close to a normal distribution. In our encompassing approach which indexes local extrema by a univariate $t$, a single normal approximation is unlikely to hold. Instead, we show that the posterior distribution of $t$ converges to a mixture of Gaussians. This multi-modal limiting distribution along with a derived local asymptotic property delineate key differences between the adopted encompassing approach and existing semiparametric work, and provide support for posterior summary that accounts for such multi-modality.

\subsection{Non-asymptotic analysis of key nonparametric quantities} \label{sec:non.asymptotic} 

In this section, we derive non-asymptotic error bounds for nonparametric recipes in Proposition~\ref{lem:closed-form}: $\widehat{\mu}_{f'}(\cdot)$, $\widehat{\sigma}^2_{f'}(\cdot)$, and their high-order derivatives under the supremum norm. These error bounds are needed to study large sample behavior of $\pi_n(t \mid X, \by)$, and might be of interest in their own right.

To this end, we take an operator-theoretic approach and make extensive use of differentiable kernels and associated properties. We begin with introducing notation and reviewing a few well-known properties; see \cite{wahba1990spline,cucker2007learning} for details. For any $f\in\Ltwo$, define the following integral operator
$
L_K(f)(x) = \int_{\mX} K(x, x') f(x') d\PP_X(x'),
$
where $x \in \mX$.
The integral operator $L_K$ is compact, positive definite, and self-adjoint. The spectral theorem ensures the existence of countable pairs of eigenvalues and eigenfunctions $(\mu_i,\psi_i)_{i\in\mathbb{N}}\subset (0,\infty)\times\Ltwo$ of $L_K$ such that $L_K\psi_i=\mu_i\psi_i$, for $i \geq 1,$ where $\{\psi_i\}_{i=1}^{\infty}$ form an orthonormal basis of $\Ltwo$ and $\mu_1\geq\mu_2\geq \cdots> 0$ with $\lim\limits_{i\rightarrow\infty}\mu_i=0$.

By Moore-Aronszajn Theorem, there is a unique reproducing kernel Hilbert space (RKHS) $\bbH$ on $\mX$ for which the Mercer kernel $K$ is the reproducing kernel. This RKHS can be characterized by a series representation $\bbH=\left\{f\in \Ltwo: \|f\|^2_{\bbH}=\sum_{i=1}^{\infty}{f_i^2}/{\mu_i}<\infty, f_i=\left<f,\psi_i\right>_{2}\right\},$ equipped with the inner product $\left<f,g\right>_{\bbH}=\sum_{i=1}^{\infty}{f_ig_i}/{\mu_i}$ for any $f=\sum_{i=1}^{\infty}f_i\psi_i$ and $g=\sum_{i=1}^{\infty}g_i\psi_i$ in $\bbH$.

We consider a proximate function of $f_0$ in $\bbH$, defined as
\begin{equation}\label{eq:def.f.lambda}
f_{\lambda} = (L_K + \lambda I)^{-1}L_K f_{0}=\sum_{i=1}^{\infty}\frac{\mu_i}{\mu_i+\la}f_i\psi_i,
\end{equation}
where $I$ is the identity operator.

We make some differentiability assumptions on $K(\cdot,\cdot)$:

\textbf{Assumption B1.} $K(\cdot,\cdot)\in C^{8}(\mX,\mX)$, i.e., $K_{jl}(x,x')=\frac{\partial ^{j+l}K(x,x')}{\partial x^{j}\partial x'^{l}}\in C(\mX,\mX)$ for any $j,l\in\mathbb{N}_0$ and $j+l\leq 8$.

Define $\kappa_{jj}=\sup_{x\in \mX}K_{jj}(x,x) > 0$ for $j=0,\ldots, 4$ and write $\kappa=\kappa_{00}$. We also define $\kappa_{0j}=\sup_{x,x'\in \mX}|K_{0j}(x,x')|$ for $j=1,\ldots,4$. Under Assumption B1, a direct application of Theorem 4.7 in~\cite{Ferreira2012} gives that $f\in C^3(\mX)$ for any $f\in\bbH$, and $\|f^{(3)}\|_\infty\leq \sqrt{\kappa_{33}}  \|f\|_\bbH$. In particular, we have $f_\la\in C^4(\mX)$ under Assumption B1. 

We further define $k$th order derivatives for $\widehat{\mu}_{f'}(x)$ and $\widehat{\sigma}^2_{f'}(x)$ as follows for $k = 0, 1, 2, 3$, with $k = 0$ corresponding to the original functions: 
\begin{align}
\label{eq:mu.higher.deriv}\widehat{\mu}^{(k)}_{f'}(x)& :=\frac{d^k}{d x^k} K_{10}(x, X) [K(X, X) + n \lambda \bI_n]^{-1} \by=K_{k+1,0}(x, X) [K(X, X) + n \lambda \bI_n]^{-1} \by, \\
\widehat{\sigma}^{2(k)}_{f'}(x) &:= \frac{d^k}{d x^k} \sigma^2 (n \lambda)^{-1} \left\{K_{11}(x, x) - K_{10}(x, X)[K(X, X) + n \lambda \bI_n]^{-1} K_{01}(X, x)\right\}\\
\label{eq:sigma.higher.deriv}&= \sigma^2 (n \lambda)^{-1}\sum_{i=0}^{k}{k\choose i}\left\{K_{i+1,k+1-i}(x,x)\right.\\ &\qquad\qquad\left. -K_{i+1,0}(x, X)[K(X, X) + n \lambda \bI_n]^{-1} K_{0,k+1-i}(X, x)\right\},
\end{align}
where \eqref{eq:sigma.higher.deriv} uses the general Leibniz rule for matrix operation.

The following Lemma~\ref{thm:nonasy.bound} establish a range of non-asymptotic error bounds under a high probability event. Let $K_{jl,x}(\cdot) = K_{jl}(x, \cdot)$ and $\varphi_{jl}(x)=(L_K+\la I)^{-1}K_{jl,x}(x)$.
\begin{lem}\label{thm:nonasy.bound}
Under Assumption B1, the following bounds for $k=0,1,2,3$ holds simultaneously under a high probability event $A_n$ with $\PP_0(A_n)\geq 1-n^{-10}$:
\begin{align}
\label{eq:higher.dev.mu.hat}&\|\widehat{\mu}^{(k)}_{f'} - f^{(k+1)}_{\lambda}\|_{\infty} \leq \frac{\sqrt{\kappa \kappa_{k+1,k+1}} \|f_0\|_\infty \sqrt{10\log n + 5}}{\sqrt{n}\lambda}  \left(10 + \frac{4 \kappa\sqrt{10\log n + 5}}{3 \sqrt{n\lambda}} \right)\\
&\qquad \qquad \qquad \qquad +\frac{C_2 \sqrt{\kappa \kappa_{k+1,k+1}} \sigma \sqrt{10\log n + 4}}{\sqrt{n} \lambda},\\
\label{eq:dev.sigma.hat.sigma}|&\widehat{\sigma}^{2(k)}_{f'}(x) - \sigma^2n^{-1} \sum_{i=0}^{k}{k\choose i} \varphi_{i+1,k+1-i}(x) |\\
&\qquad \leq\sum_{i=0}^{k}{k\choose i}\left[\frac{\sqrt{\kappa \kappa_{i+1,i+1}} \kappa_{0,k+1-i}\sigma^2 \sqrt{10\log n+4}}{n\sqrt{n}\lambda^2}  \left(10 + \frac{4 \sqrt{\kappa}\sqrt{10\log n+4}}{3 \sqrt{n\lambda}} \right)\right].
\end{align}
\end{lem}

The following assumption allows us to simplify the bounds for  $\widehat{\sigma}^{2(k)}_{f'}(x)$ for $k = 0, 1, 2, 3$. 

\textbf{Assumption B2.} $\la \sup_{x\in \mX}|\varphi_{jl}(x)|$ is bounded for $j,l\geq 1$ and $j+l\leq 5$.

\begin{remark}
Under Assumption B2, Equation~\eqref{eq:dev.sigma.hat.sigma} yields
$\|  \widehat{\sigma}^2_{f'}(x) -  \sigma^2n^{-1}\varphi_{11}(x) \|_{\infty} \lesssim \frac{ \sqrt{\log n}}{n\sqrt{n}\lambda^2}$ and $
\|\widehat{\sigma}^{2(k)}_{f'}(x) \|_\infty \lesssim \frac{1}{n\la}$ for $k = 1, 2, 3.$ Hence, $\widehat{\sigma}^2_{f'}(x)$ approximates $\varphi_{11}(x) \sigma^2/n$ with high probability.
\end{remark}

We make the following assumption to ease the presentation. The subsequent theory in Theorems~\ref{thm:LAN} and~\ref{thm:bvm} can be generalized for cases where Assumption B3 does not hold, with more complicated expressions that involve $K_{11}(x, x)$ and its derivatives. 

\textbf{Assumption B3.} $K_{11}(x,x)$ does not depend on $x$.

Assumption B3 simplifies the posterior $\pi_n(t\mid X,\by)$ in Proposition~\ref{lem:closed-form}. It holds for any stationary kernels; this is because if $K(x, x') = g(x - x')$ for some function $g(\cdot)$, then $K_{11}(x, x) = -g''(0)$, which does not depend on $x$. 

The following assumption is concerned with the error term $f'_\la-f'_0$. This is a deterministic function as $f_\la$ does not depend on random draws of covariates and noise. 

\textbf{Assumption C.} $\|f'_\lambda-f'_0\|_\infty\lesssim \la^{r_1}$ and $\|f''_\lambda-f''_0\|_\infty\lesssim \la^{r_2}$ for some $0< r_1,r_2\leq 1$.

Assumption C ensures that $f_\la'$ and $f''_\la$ converge to $f'_0$ and $f''_0$ under the supremum norm, respectively. The two parameters $r_1$ and $r_2$ correspond to approximation properties of $f_{\lambda}$ to the function class that $f_0$ belongs to, and such properties in turn depend on the covariance kernel and smoothness of the function class.
Assumption C is typically verifiable via direct calculation for a given problem. For example, if $L_K^{-r'-\frac{1}{2}}f_0\in \Ltwo$ for some $0<r'\leq \frac{1}{2}$, then the one-dimensional case of Theorem 6 in \cite{liu2023estimation} gives $r_1 = r_2 = r'$; here the function class for $f_0$ is defined by integral operators. 

Assumptions B1-B3 spell out conditions for the kernel and regularization parameter $\lambda$, while Assumption C is a generic condition that includes a range of function classes of $f_0$. In Section~\ref{sec:determinstic.bound} we provide examples where Assumptions B1-B3 and C hold.

\subsection{LAN property}\label{sub:LAN}

The following Lemma~\ref{lem:post.mode} shows existence of local extrema of $f_\la$ and $\widehat{\mu}_f$ within a neighborhood of $t_m$, which are respectively denoted by $t_{\la, m}$ and $\hat{t}_m$ for $m = 1, \ldots, M$.

\begin{lem}\label{lem:post.mode}
Under Assumptions A1-A3, B1 and C, for any sufficiently small $\la$, there exist $\{t_{\la,m}: m = 1, \ldots, M\}$ such that each $t_{\la,m}$ is a local extremum of $f_\lambda$ and $|t_{\la,m}-t_m| \lesssim \la^{r_1}.$ Moreover, under $A_n$, there exist $\{\hat t_m: m = 1, \ldots, M\}$ such that each $\hat t_m$ is a local extremum of $\widehat{\mu}_{f}$ and $|\hat t_m-t_{\la,m}| \lesssim {\sqrt{\log n}}/{(\sqrt n \la)}.$
\end{lem}

Henceforth we work under the high probability event $A_n$ defined in Lemma~\ref{thm:nonasy.bound}. Let the regularization parameter $\la=n^{-\frac{1}{2}+\beta}(\log n)^{\frac 12+a}$ for some $0<\beta<\frac12$ and $a>0$. Since $f'_\la(t_{\la,m})=0$, Lemma~\ref{thm:nonasy.bound} implies that 
\begin{align}
\label{eq:mu.hat}|n^\beta\widehat{\mu}_{f'}(t_{\la,m})|&\lesssim (\log n)^{-a},\\
\label{eq:mu.hat.prime} |\widehat{\mu}_{f'}^{(k)}(t_{\la,m})-f_\la^{(k+1)}(t_{\la,m})| & \lesssim n^{-\beta}(\log n)^{-a} ,\quad 1\leq k\leq 3, \\
\label{eq:sigma.t.hat}
\bigg|n\widehat{\sigma}^2_{f'}(t_{\la,m})-\sigma^2\varphi_{11}(t_{\la,m})\bigg| & \lesssim {n^{\frac{1}{2}-2\beta}}(\log n)^{-1-a}, \\
|\widehat{\sigma}_{f'}^{2(k)}(t_{\la,m})| & \lesssim {n^{-\frac{1}{2}-\beta}}(\log n)^{-\frac 12-a},\quad 1\leq k\leq 3. \label{eq:higher.sigma.t.hat}
\end{align}

The following Theorem~\ref{thm:LAN} characterizes the marginal likelihood function of $t$ by presenting a local asymptotic normality (LAN) property at $t_{\la,m}$, generalizing traditional LAN properties to the considered encompassing semiparametric regime. For $m = 1, \ldots, M$, we denote by
\begin{equation}\label{eq:def.varphi.sigma.m}
\varphi_{11,m}=\varphi_{11}(t_{\la,m}), \quad \sigma_m^{*2}={\sigma^2}/{f_0''(t_m)^2}.
\end{equation}
\begin{thm}\label{thm:LAN}
Suppose Assumptions A1-A3, B1-B3 and C hold. Let $\la=n^{-\frac{1}{2}+\beta}(\log n)^{\frac 12 + a}$ for some $\frac 14<\beta<\frac12$ and $a>0$. Suppose $n^{\frac{3}{2}-4\beta}\varphi^{-2}_{11,m}$ is bounded for $m=1,\ldots,M$. Then under event $A_n$ as $n\rightarrow \infty$, for any $m=1,\ldots,M$, the marginal likelihood likelihood $\ell(t)$ satisfies the LAN property
\begin{equation}\label{eq:LAN}
\log\frac{\ell(t_{\la,m}+\frac{u}{n^\beta})}{\ell(t_{\la,m})}=n^{1-2\beta}\varphi_{11,m}^{-1}\left(-\frac{u^2}{2\sigma_{n,m}^2}-\frac{\mu_{n,m}u}{\sigma_{n,m}^{2}}\right)+o(1),
\end{equation}
where $\mu_{n,m}=\frac{n^\beta\widehat{\mu}_{f'}(t_{\la,m})}{\widehat{\mu}_{f'}'(t_{\la,m})}$ and $\sigma^2_{n,m}=\frac{n\varphi_{11,m}^{-1}\widehat{\sigma}^2_{f'}(t_{\la,m})}{\widehat{\mu}_{f'}'(t_{\la,m})^2}$. 

Moreover, letting $r=r_1 \wedge r_2$, we have
\begin{equation}\label{eq:mu.ni}
|\mu_{n,m}|\lesssim (\log n)^{-a},
\end{equation}
\begin{equation}\label{eq:sigma.ni}
|\sigma^2_{n,m} -\sigma_m^{*2}| \lesssim \la^r.
\end{equation}
\end{thm}
\begin{remark}
The LAN property in Theorem~\ref{thm:LAN} exhibits two differences compared to classical LAN properties (Section 7, \cite{van2000asymptotic}, \cite{kleijn2012bernstein}), in part owing to the adopted unconventional encompassing approach along with the semiparametric problem under consideration. First, the inflation term $n^{1-2\beta}\varphi_{11,m}^{-1}$ is absent in classical semiparametric LAN expansions, pointing to complications in the rate calculation that integrate properties of $K$ and the choice of $\lambda$ through $\varphi_{11,m}$. Second, our expansion is specific to $t_{\la, m}$, a local extremum of $f_{\la}$, with varying quantities $\mu_{n, m}$ and $\sigma^2_{n, m}$. As a result, the posterior distribution cannot be approximated by a Gaussian after a homogeneous rescaling of the parameter across $\mX$; this reminds us of the \textit{localized} feature of local extrema, and suggests localized rescaling within a small interval centered at $t_m$. Unlike existing work in semiparametric Bayes where the posterior distribution converges to a single Gaussian, a multi-modal posterior distribution in the form of a mixture of normal distributions is expected.
\end{remark}

\subsection{Multi-modal limiting distribution} \label{sec:BvM}

We make the following mild assumption on the prior $\pi(t)$:

\textbf{Assumption D.} The prior density $\pi(t)$ satisfies that $\pi(t)\in C(\mX)$ and has positive density at local extrema, i.e., there holds $\pi(t_m)> 0$ for $m=1,\ldots,M$.

This assumption on $\pi(t)$ is rather flexible and can be satisfied by most continuous distributions supported on $\mX$, such as the beta distribution.

Let $\Pi_n(\cdot\mid  X,\by)$ be the probability measure of $\pi_n(\cdot\mid  X,\by)$, and $\Phi(\cdot \mid  \mu, \sigma^2)$ be the cumulative distribution function of the normal distribution $N(\mu, \sigma^2)$. The following theorem shows that $\pi_n(t\mid X,\by)$ is close to a mixture of normal distributions when the sample size is large, where the component densities are related to the LAN expansion, and the weights are determined by both prior density and curvature of $f_0$ at $t_m$.

\newcommand{\rate}{\epsilon_n}
\begin{thm}\label{thm:bvm}
Suppose Assumptions A1-A3, B1-B3, C and D hold, and let $\la=n^{-\frac{1}{2}+\beta}(\log n)^{\frac 12 + a}$ for some $\frac 14<\beta<\frac12$ and $a>0$ such that $n^{\frac{3}{2}-4\beta}\varphi^{-2}_{11,m}$ is bounded. Then the following results hold for any $z \in \RR$:

(i) Letting $\pi_m=\frac{|f_0''(t_m)|^{-1}\pi(t_m)}{\sum_{m=1}^{M}|f_0''(t_m)|^{-1}\pi(t_m)}$, we have 
\begin{equation} \label{eq:bvm.unrescaled}
\bigg|\Pi_n(t\leq z\mid X,\bm y) - \sum_{m=1}^{M}\pi_m\Phi(z\mid t_m,n^{-1}\varphi_{11,m}\sigma_m^{*2})\bigg| \rightarrow 0
\end{equation}
in $\PP_0$-probability, where $\varphi_{11,m}$ and $\sigma_m^{*2}$ are given by~\eqref{eq:def.varphi.sigma.m}.

(ii) Letting $\Pi_{n,m}'(\cdot\mid X,\by)$ be the posterior of $\sqrt{\frac{n}{\varphi_{11,m}}}(t \ind_{I_m}(t) -\hat t_m+b_n)$ where $b_n= n^{-\beta} \log n$, and $I_m=[t_m-\zeta_{m-1},t_m+\zeta_m]$ with $\zeta_0=t_1$, $\zeta_m=(t_{m+1}-t_{m})/2$ for $m=1,\ldots, M-1$, and $\zeta_M=1-t_M$, we have for any $z \in \RR$, in $\PP_0$-probability, 
\begin{equation} \label{eq:bvm.rescaled}
\bigg|\Pi_{n,m}'(t'\leq z\mid X,\bm y) - \Phi(z\mid 0,\sigma_m^{*2})\bigg|\rightarrow 0.
\end{equation}
\end{thm}
\color{black}

A few remarks are in order to elucidate the encompassing strategy using Theorem~\ref{thm:bvm}.

\begin{remark}
Part (i) of Theorem~\ref{thm:bvm} shows that the posterior distribution is close to a mixture of normal distributions. The curvature of $f$ at a local extremum $t$, defined as ${|f''(t)|}/(1+\{f'(t)\}^2)^{3/2}$ that reduces to $|f''(t)|$ when $f'(t) = 0$, directly affects the variation of the posterior distribution at $t$. Indeed, the standard derivation of the Gaussian component at $t$ in the limiting distribution is proportional to $1/|f_0''(t)|$, meaning a large curvature leads to a more concentrated normal component in the posterior distribution. Interestingly, this effect of curvature on the component-wise variance is offset by the mixture weight that is also proportional to $1/|f_0''(t)|$. More specifically, the limiting mixture normal density function evaluated at each local extremum $t_m$, which is $\pi_m /\sqrt{2 \pi n^{-1}\varphi_{11,m}\sigma_m^{*2}} \propto \pi(t_m)/\sqrt{\varphi_{11,m}},$ does not depend on the curvature of $f$ at $t_m$. Hence, the multi-modal posterior distribution of $t$ does not diminish a local extremum with small curvature, at least asymptotically. The prior weight $\pi(t_m)$ plays a direct role in driving the posterior distribution, which enables incorporating prior knowledge. Section~\ref{sec:simulation.one.replicate} provides numerical confirmation for these theoretical implications using finite sample illustration; see, in particular, Figure~\ref{fig:effect.n}. The marginal likelihood function $\ell(t)$, proportional to the posterior density $\pi_n(t \mid X, \by)$ with a uniform prior, 
also tends to be multi-modal, as observed in Figure~\ref{fig:effect.n}.
\end{remark}

\begin{remark}
Part (ii) generalizes the BvM phenomenon to the encompassing semiparametric regime under consideration. In particular, after rescaling and truncation, the posterior distribution weakly converges to a normal distribution with a bias term $b_n$ that is $o(1).$
BvM theorems in weak convergence (i.e., convergence in distribution) are common in the literature \citep{kim2004bernstein,castillo2015bernstein,castillo2014bernstein,kim2006bernstein}. Our result is different in that the target distribution is  multi-modal with varying mean and variance at each component, necessitating a localized truncation at $I_m$ for weak convergence to a single normal. Although not typical, approximating the posterior distribution via a mixture of Gaussians has appeared in the literature; for example, see \cite{Castillo2015linearmodel} on Bayesian linear regression models. The established results and proofs show other important differences from the semiparametric literature. \cite{castillo2012semiparametric} proposed sufficient conditions for a BvM theorem for separated models, where the model parameter takes the form $\eta=(\theta,f)$. In our case, $t$ is an inherited hyperparameter of $f$ rather than an independent parameter in a separated model (e.g., parameters in a location-scale family). \cite{castillo2015bernstein} provided sufficient conditions for a BvM theorem for smooth functionals of the parameter in general models. Specifically, they considered a model parameterized by $\theta\in\Theta$, and provided a BvM theorem for $\psi(\theta)$ where $\psi :\Theta\rightarrow \RR^d$ is a smooth functional of interest (see Equation (2.4) in their paper). However, the derivative at any fixed point, when viewed as a functional of the regression function, is not bounded~\citep[page 13]{conway1994course}, and thus local extrema may not be expressed as a functional of the regression function even when their number is known.
\end{remark}

\begin{remark}
Part (ii) indicates a bias-variance trade-off regarding the rescaled posterior for estimating $t_m$. The bias of the centering quantity $\hat t_m-b_n - t_m$ is bounded above by $\lambda^{r_1}$, in view of Lemma~\ref{lem:post.mode} and the constraint that $\frac{1}{4} < \beta < \frac{1}{2}$. On the other hand, the variance $n^{-1}\varphi_{11,m}$ is bounded above by $(n\lambda)^{-1}$ in view of Assumption B2. Therefore, a larger $\lambda$ (or equivalently, a larger $\beta$) corresponds to a smaller variance but a larger bias. An exact rate calculation for both bias and variance can be obtained by considering the special cases in Section~\ref{sec:determinstic.bound}. Note that such rates depend on the underlying regularity parameter of the true function that is typically unknown, impeding their application to parameter tuning. We propose to use an empirical Bayes approach to select $\lambda$ by maximizing its marginal likelihood function, which shows competitive performance in our simulations; see Section~\ref{sec:simulation} for details. 
\end{remark}

Verifying the conditions needed for the preceding theorems often amounts to checking Assumptions B1-B3 and Assumption C, which will provide insight into how to choose the kernel hyperparameters; see Section~\ref{sec:determinstic.bound} for examples.

\subsection{Point and interval estimation} \label{sec:point.interval.est} 
The shape of the posterior distribution characterized in Theorem~\ref{thm:bvm} can be used to construct estimators with frequentist properties. In particular, the multi-modality of the limiting posterior distribution provides a basis to overcome the multiplicity challenge of local extrema, and leads to consistent estimators of $t_m$ through posterior exploration. We additionally provide interval estimation that achieves frequentist coverage. 
\begin{thm} \label{thm:frequentist}
Under the same conditions as in Theorem~\ref{thm:bvm}, the following results hold. 

(i) For all $\delta > 0$, with $\PP_0$-probability tending to one, $\pi_n(t \mid X,\bm{y})$ has exactly $M$ local maxima $t_{1,n}, \ldots, t_{M,n}$, and for $m=1,\ldots, M$ there holds $|t_{m,n} - t_m | \leq \delta$ for all sufficiently large $n$. 

(ii) Let $\Delta_{n}(\cdot) = K_{10}(\cdot, X) [K(X, X) + n \lambda \bm I_n]^{-1} f_0(X)$. 
For any $\alpha \in (0, 1)$, the following is an asymptotic $1-\alpha$ confidence interval for $t_m + \Delta_n(t_m)/f_0''(t_m)$: 
\begin{equation}
\hat t_m \pm z_{\alpha/2} { \sigma \sqrt{K_{10}(\hat t_m, X) [K(X, X) + n \lambda \bm I_n]^{-2} K_{10}(\hat t_m, X)^T}}{ /|\widehat\mu_{f'}'(\hat t_m)| },
\end{equation}
where $z_{\alpha/2}$ is the upper $\alpha/2$ quantile of the standard normal distribution. 
\end{thm}
In Theorem~\ref{thm:frequentist} (ii) the bias term $\Delta_n(t_m)/f_0''(t_m)$ can be shown to be $o(1)$ with $\PP_0$-probability tending to one as it is approximating $f'_0(t_m)/f_0''(t_m) = 0$. 
For implementation, we propose to ``plug in'' consistent estimators for unknown quantities in $\Delta_n(t_m)/f_0''(t_m)$, including $\widehat\mu_{f'}'(\hat t_m)$ for $f_0''(t_m)$, $\hat{t}_m$ for $t_m$, and $\widehat\Delta_{n}(\cdot) = K_{10}(\cdot, X) [K(X, X) + n \lambda \bm I_n]^{-1} \widehat\mu_f(X)$ for $\Delta_n(\cdot)$, leading to the following confidence intervals for $t_m$:
\begin{equation} \label{eq:ci}
\hat t_m + \frac{\widehat\Delta_n(\hat t_m)}{\widehat{\mu}'_{f'}(\hat t_m)} \pm z_{\alpha/2} \frac{ \sigma \sqrt{K_{10}(\hat t_m, X) [K(X, X) + n \lambda \bm I_n]^{-2} K_{10}(\hat t_m, X)^T}}{ |\widehat\mu_{f'}'(\hat t_m)| }.
\end{equation}
These confidence intervals depend on $\lambda$, which will be estimated using empirical Bayes, as discussed in Remark 5. In the context of nonparametric regression, \cite{liu2022optimal} have demonstrated that this choice of $\lambda$ tends to adapt to the unknown smoothness level of the underlying function, especially when combined with an oversmooth kernel. However, it is worth noting that constructing adaptive confidence intervals with minimax optimal diameter is a more challenging task, as is typically the case in nonparametric inference; see, for example, \cite{gine2021mathematical} for more details. We assess finite sample performance of~\eqref{eq:ci} in Section~\ref{sec:simulation}, which shows satisfactory coverage.

When the error variance $\sigma^2$ is unknown, one may substitute $\sigma^2$ in the original derivative-constraint GP prior with an estimate $\hat{\sigma}^2_n$. 
The established results in Theorems~\ref{thm:LAN},~\ref{thm:bvm}, and~\ref{thm:frequentist} hold with any estimator $\hat{\sigma}^2_n$ that converges to $\sigma^2$ in mean square. In particular, we can estimate $\sigma^2$ by the maximum marginal likelihood estimator which has been shown to be mean square consistent under various settings~\citep{yoo2016supremum,liu2022optimal}. Denote the induced posterior measure of $t$ by $\Pi_{n,{\hat{\sigma}^2_n}}(\cdot\mid X,\by)$, and take Part (i) in Theorem~\ref{thm:bvm} as an example.
Let $\mathcal{B}_n$ be a shrinking neighborhood of $\sigma^2$ such that $\PP_0(\hat{\sigma}^2_n\in\mathcal{B}_n)\rightarrow 1$. Conditional on $\mathcal{B}_n$, Equation~\eqref{eq:dev.sigma.hat.sigma} becomes 
\begin{align}
&|\widehat{\sigma}^{2(k)}_{f'}(x) - (\sigma^2+o(1))n^{-1} \sum_{i=0}^{k}{k\choose i} \varphi_{i+1,k+1-i}(x) |\\ &\qquad \leq\sum_{i=0}^{k}{k\choose i}\left[\frac{\sqrt{\kappa \kappa_{i+1,i+1}} \kappa_{0,k+1-i}(\sigma^2+o(1))\sqrt{10\log n+4}}{n\sqrt{n}\lambda^2}  \left(10 + \frac{4 \sqrt{\kappa}\sqrt{10\log n+4}}{3 \sqrt{n\lambda}} \right)\right],
\end{align}
and all the established inequalities in the proof of Theorem~\ref{thm:bvm} hold uniformly over $\sigma^2\in \mathcal{B}_n$. In particular, there holds
$
\sup_{\sigma^2\in\mathcal{B}_n}\bigg|\Pi_n(t\leq z\mid X,\bm y) - \sum_{m=1}^{M}\pi_m\Phi(z\mid t_m,n^{-1}\varphi_{11,m}\sigma_m^{*2})\bigg| \rightarrow 0
$
in $\PP_0$-probability, yielding $\bigg|\Pi_{n,\hat\sigma^2_n}(t\leq z\mid X,\bm y) - \sum_{m=1}^{M}\pi_m\Phi(z\mid t_m,n^{-1}\varphi_{11,m}\sigma_m^{*2})\bigg| \rightarrow 0.$

\subsection{Applications to special function classes and GP kernels} \label{sec:determinstic.bound}

In this section, we provide examples under which various assumptions and Theorems~\ref{thm:LAN} and~\ref{thm:bvm} hold. We focus on covariance kernels that possess regularized eigenfunctions as follows.

\textbf{Assumption E.} The eigenfunction $\psi_i\in C^s(\mX)$ for all $i\in\mathbb{N}$ and some $s\in\mathbb{N}$. Moreover, there exists a constant $C>0$ such that $\|\psi_i^{(s)}\|_\infty\leq Ci^s$ for any $i\in\mathbb{N}$ and $\s \psi_i'(x)^2$ diverges for any $x \in \mX$. 

In particular, the Fourier basis satisfies Assumption E for any finite $s$. 

\textit{Example 1. Stationary kernels with polynomially decaying eigenvalues. } Let $K^{\alpha,s}$ be a stationary covariance kernel whose eigenfunctions satisfy Assumption E and eigenvalues decay at a polynomial rate, that is, 
$\mu_i\asymp i^{-2\alpha}
$
for $i\in\mathbb{N}$ and some $\alpha>0$.

We assume that the true regression function $f_0$ lies in the H\"older class:
\begin{equation}
H^{\alpha}(\mX)=\left\{f\in \Ltwo: \|f\|^2_{H^{\alpha}(\mX)}=\sum_{i=1}^{\infty}i^{\alpha}|f_i|<\infty,f_i=\left<f,\psi_i\right>_2\right\}.
\end{equation}
Any function in $H^{\alpha}(\mX)$ has continuous derivatives up to order $\lfloor\alpha\rfloor$ and the $\lfloor\alpha\rfloor$th derivative is Lipschitz continuous of order $\alpha-\lfloor\alpha\rfloor$. Using the error bounds for derivatives of $f_\la - f_0$ in Lemma 13 of~\cite{liu2023estimation} to verify Assumption C, the following corollary provides an example for Theorems~\ref{thm:LAN} and~\ref{thm:bvm} to hold.

\begin{thm}\label{thm:example.poly}
Suppose $f_0\in H^\alpha(\mX)$ and $K^{\alpha,s}$ is used in the GP prior for $\alpha> 9/2$ and $s\geq 4$. Then Assumptions B1-B3 and C hold, and Theorems~\ref{thm:LAN} and~\ref{thm:bvm} hold for any $\beta\in [\frac 38,\frac 12)$.
\end{thm}

\textit{Example 2. Stationary kernels with exponentially decaying eigenvalues.} 
We consider stationary kernels $K_{\gamma,s}$ with eigenvalues $\mu_i\asymp e^{-2\gamma i}$ for $i\in\mathbb{N}$ and some $\gamma>0$, and eigenfunctions satisfying Assumption E. The well-known squared exponential kernel can be approximately viewed as an example of $K_{\gamma,s}$, with a closed-form eigendecomposition with respect to Gaussian sampling on the real line \citep{rasmussen2006,pati2015adaptive}.

We assume $f_0$ belongs to the analytic-type function class $A_\gamma(\mX)$:
\begin{equation}
A_\gamma(\mX)=\left\{f\in \Ltwo: \|f\|^2_{A_\gamma(\mX)}= \sum_{i=1}^{\infty}e^{\gamma i}|f_i|<\infty,f_i=\left<f,\psi_i\right>_2\right\}.
\end{equation}

We first verify Assumption C in the following Lemma~\ref{lem:exp.deriv.deterministic}.
\begin{lem}\label{lem:exp.deriv.deterministic}
Suppose that $f_0\in A_\gamma(\mX)$, and $K_{\gamma,s}$ is used in the GP prior for some $\gamma>\frac{k}{e}$ and $s\geq k$. Then there holds $\|f_\lambda^{(k)}-f_0^{(k)}\|_\infty \lesssim \la^{\frac 12 - \frac{k}{2e\gamma}}.$
\end{lem}

This yields another example for our theory to hold using kernels with exponentially decaying eigenvalues, formulated in the following corollary. 
\begin{thm} \label{thm:example.exp}
Suppose $f_0\in A_\gamma(\mX)$ and $K_{\gamma,s}$ is used in the GP prior for $\gamma>\frac{5}{2e}$ and $s\geq 4$. Then Assumptions B1-B3 and C hold, and Theorems~\ref{thm:LAN} and~\ref{thm:bvm} hold for any $\beta\in [\frac 38,\frac 12)$.
\end{thm}

\section{Simulation} \label{sec:simulation} 
In this section we carry out simulation studies to illustrate the convergence of the posterior distribution $\pi_n(t \mid X, \by)$ to Gaussian mixtures, and to assess the performance of the proposed method relative to competing methods. 

\begin{figure}
\centering
\includegraphics[width=.5\textwidth]{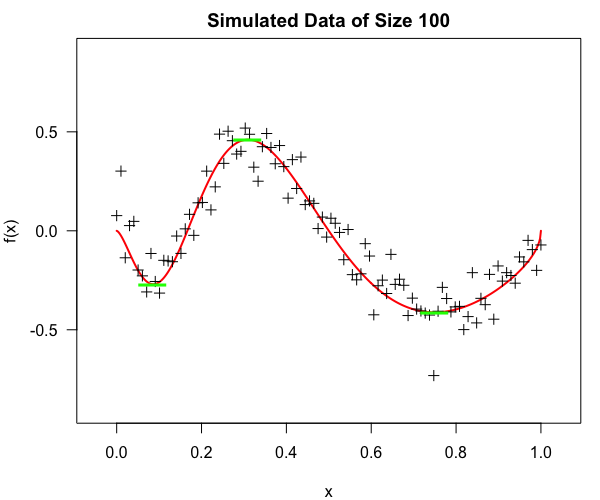}
\caption{Simulated data with $n = 100$. Observations are marked by ``+''. The red curve is the true regression function, and green lines indicate the location of three local extrema.}
\label{fig:simulated_data_100}
\end{figure}

We use a Doppler-type regression function $f(x) = \sqrt{x(1-x)}\sin \left({2\pi}/{(x+0.5)} \right)$ for $x \in \mathcal{X} = [0, 1],$ which has three local extrema at $t_1 = 0.0863$, $t_2 = 0.3096$ and $t_3 = 0.7491$. We add iid zero-mean Gaussian noise to $f$ with standard deviation $\sigma = 0.1$, observed at equal-spaced
$\{x_i\}_{i=1}^n$ in the unit interval $[0, 1]$. We vary the sample size $n = 100, 500, 1000$. Figure \ref{fig:simulated_data_100} shows one simulated dataset of sample size 100. Each simulation scenario is replicated 100 times.

We use two prior distributions Beta(1, 1) (the uniform distribution) and Beta(2,3) on $t$ to study the sensitivity of posterior inference to the prior specification. We use the squared exponential kernel function for $K$, that is, $K(x, x') = \exp\{-(x - x')^2/(2 h^2)\}$. For each simulated dataset, we select parameters other than $t$, which include $\lambda, h$, and $\sigma^2$, via an empirical Bayes approach by maximizing the unconstrained marginal likelihood function, i.e., the multivariate normal density $N(0, \sigma^2 (n \lambda)^{-1}K(X, X) + \sigma^2 \bI_n)$ evaluated at $\by$. This is motivated by the excellent performance of empirical Bayes in a variety of settings \citep{yoo2016supremum,liu2022optimal}. We use the midpoint rule to calculate the normalizing constant in $\pi_n(t \mid X, \by)$. 

\subsection{Finite sample size behavior of the posterior distribution} \label{sec:simulation.one.replicate}
\begin{figure}
\centering
\begin{tabular}{cc}
\includegraphics[trim=0 0 0 35, width=0.45\linewidth, clip=TRUE]{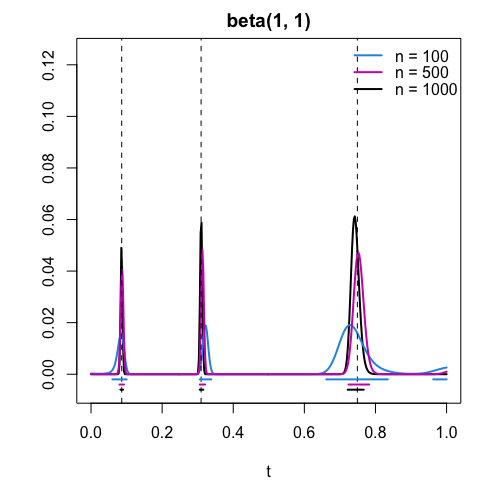}     &  
\includegraphics[trim=0 0 0 35, width=0.45\linewidth, clip=TRUE]{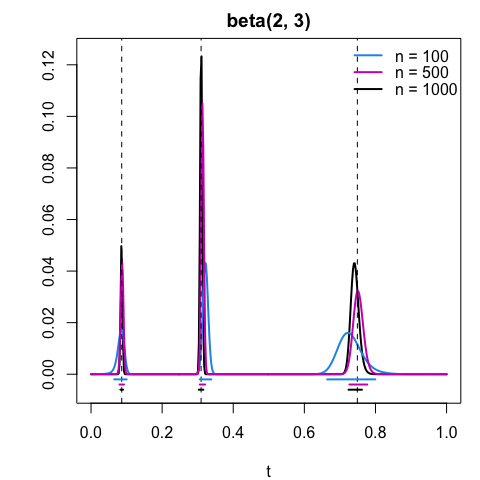} \\
(a) Beta(1, 1) prior & (b) Beta(2, 3) prior
\end{tabular}
\caption{Effect of $n$ on the posterior distribution of $t$ with beta prior distributions. Vertical dashed lines indicate the true locations of local extrema. The intervals in each plot are the 95\% highest posterior density regions. Each posterior density function is based on one simulated dataset.} \label{fig:effect.n}
\end{figure}

The proposed method, labeled as DGP, does not require any sampling to obtain the posterior distribution owing to the closed-form expression in Proposition~\ref{lem:closed-form}. Figure \ref{fig:effect.n} shows the posterior distribution of $t$ with various sample sizes and the two beta priors, each based on one simulated dataset. We can see that the posterior distribution possesses three mixture components at all sample sizes and for both beta priors, matching the true number of local extrema $M = 3$. At each sample size such as $n = 1000$, the posterior distribution tends to have three modes concentrating around $(t_1, t_2, t_3)$, which aligns with the established Theorem~\ref{thm:bvm} that deciphers the limiting behavior of the posterior distribution. 

The curvature at a local extremum point $t$ is $|f''(t)|$, which is $(111.04, 44.55, 11.91)$ for $(t_1, t_2, t_3)$, respectively. Figure~\ref{fig:effect.n} indicates that the variability of each mixture component decreases substantially as the curvature increases, confirming Theorem~\ref{thm:bvm} in which we show that the standard deviation of each Gaussian component in the limiting distribution is inversely proportional to the curvature at the corresponding local extremum. For example, $t_1$ with the highest curvature exhibits the least variation, while $t_3$ with the lowest curvature has the most variation in the posterior distribution, as in Figure~\ref{fig:effect.n}.

As $n$ increases, the mixture components in the posterior distribution are more bell-shaped. When the sample size is small, such as $n = 100$, the mixture component may be skewed. This is particularly the case for the first (left skewed) and the third mixture component (right skewed) when the Beta(1,1) prior is used. A closer inspection of Figure~\ref{fig:effect.n} (a) indicates a \textit{boundary effect} when $n = 100$, that is, there appears to be a small bump near the boundary. Such boundary effects are reasonable as there are sparser data near the boundary, lacking information outside the range $\mathcal{X}$, and that the right boundary point $t = 1$ indeed gives the largest function value on $(t_3, 1]$. Both skewed mixture components and boundary effects are much mitigated when the sample size increases to 500 and 1000. In addition, the Beta(2, 3) prior tends to zero out the boundary bumps even when the sample size is as small as 100. The posterior distributions corresponding to the two beta priors have various density values at their local peaks, which are also suggested by Theorem~\ref{thm:bvm}. We remark that, however, an appropriate posterior summary method such as highest posterior density regions may lead to interval estimates that are less sensitive to the priors; see the interval estimates in Figure~\ref{fig:effect.n}, and the next section, particularly Table~\ref{table:RMSE}, for more details about point estimates. 

The proposed Bayesian approach has the advantage of allowing users to incorporate any prior knowledge, if available, about the location of local extrema. For example, one may use a suitable prior distribution to rule out the possibilities of local extrema near the boundary. This does not necessarily mean that local extrema are not located near the boundary, but rather that such local extrema are not desired. 

\subsection{Comparison with other methods}

We use the 95\% highest posterior density region (HPDR) of the posterior distribution of $t$ for posterior summary, which consists of a number of disjoint intervals enclosing local modes. We use the number of segments in the HPDR to estimate $M$, and the corresponding posterior mode within each segment to estimate local extrema. This posterior summary is a reasonable strategy according to our asymptotic characterization of the posterior distribution of $t$, which approximates a mixture of Gaussians with the number and location of the mixture components matching the local extrema of the underlying regression function. We expect this method to estimate $M$ correctly with high probability according to Theorem~\ref{thm:frequentist} (i).

For comparison, we implement another three methods: the smoothed taut string (STS) method proposed by~\cite{Kovac2007}, the original taut string (TS) method in~\cite{davies2001local}, and the nonparametric kernel smoothing (NKS) method proposed by~\cite{Song2006}. STS and TS estimate the number of local extrema by minimizing a loss function of the corresponding taut string, and do not provide uncertainty quantification about local extrema. Since STS is an improved version of TS, and we find that these two methods lead to similar numerical performance in our experiments, in this section we omit the results of TS and use STS to represent taut string-based methods. NKS first estimates the regression function, denoted by $\hat{f}(x)$, then chooses a set of $x$'s such that the confidence interval of $f'(x)$ contains zero. Within this set, point estimates of stationary points are obtained by locating those at which $\hat{f}'(x)$ are closest to zero, denoted by $\{x_m^*\} _{m=1}^{\hat{M}}$, which are a subset of $\{x_i\}_{i=1}^n$ by design. For interval estimation, NKS inverts the lower and upper limits of the 95\% confidence band for $\hat{f}'(x_m^*)$. Such intervals may not exist, and if one of the upper or lower bounds can be found, they further assume asymptotic normality and construct a symmetric interval based on $x_m^*$ and the available bound. In contrast, interval estimation in the semiparametric Bayesian approach through HPDRs is computationally more  straightforward and conceptually more coherent. STS and TS are implemented in the R package \texttt{ftnonpar}, and we implement NKS using the R code provided by the authors of \cite{Song2006}. The bandwidth parameter in NKS is chosen by minimizing asymptotic mean integrated squared error.

\subsubsection{Estimation of $M$} \label{sec:sim.M}

\begin{figure}[hbt!]
\centering
\includegraphics[width=0.415\linewidth, trim = 0 0 100 0, clip = TRUE]{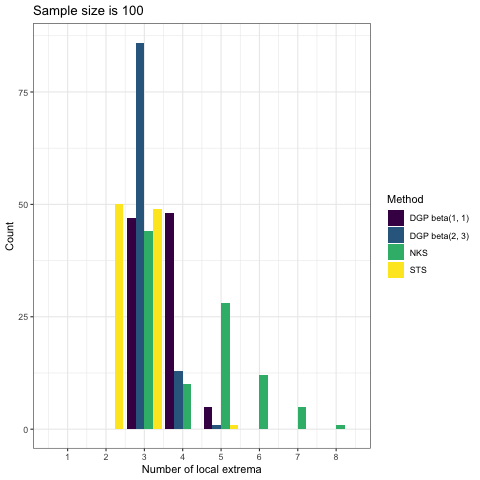}
\includegraphics[width=0.52\linewidth]{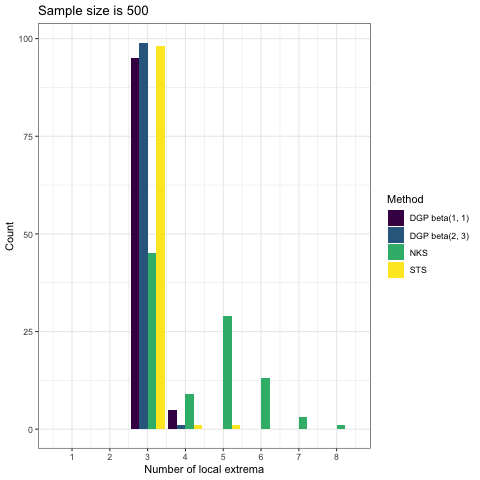}
\caption{Frequency of the estimated number of local extrema by each method across 100 replicated simulations. Color code: yellow for STS, green for NKS, blue for DGP with the Beta(2,3) prior, and purple for DGP with the Beta(1, 1) prior. The plot with $n = 1000$ is similar to the one with $n = 500$, and is thus omitted here.} \label{fig:freq}
\end{figure}

Figure~\ref{fig:freq} plots the estimated number of local extrema by each method at various sample sizes. The plot with $n = 1000$ is similar to the one with $n = 500$, and is thus omitted here. We can see that the semiparametric Bayesian method DGP with both priors and the STS method tend to capture the true number of local extrema as the sample size increases to 500. When the sample size is 100, DGP with the Beta(2,3) prior gives the largest frequency at the true number $M = 3$ among all methods, while STS and DGP with the Beta(1,1) prior either underestimate or overestimate $M$ in about half of the 100 simulations, although mostly by one. Figure~\ref{fig:freq} reinforces that the estimation accuracy of $M$ is greatly improved by using the Beta(2, 3) prior to remove trivial points around the boundary, which gives the most accurate estimate of $M$ for both $n = 100$ and $n = 500$. It is reassuring that the effect of prior distributions for DGP is diminished as $n$ increases from 100 to 500. For both DGP and STS, a sample size over 500 appears large enough to ensure an accurate estimate of the number of local extrema, at least under the simulation setting. NKS does not exhibit a clear convergence behavior as DGP and STS do. It overestimates $M$ considerably more than DGP and STS, and such overestimation persists when the sample size increases to 500 and 1000. This confirms that the presence of multiple local extrema poses challenges to NKS, as commented by \cite{Song2006}, and suggests that multiple testing correction is particularly needed for the two-step approach, while appealing performance of the unified approach DGP does not hinge on such a correction. 

Additional experiments are included in the supplementary material to investigate the effects of noise standard deviation and credible levels, which show quite robust performance of the Beta(2,3) prior in estimating $M$ for a wide range of credible levels. A highly fluctuated regression function with large $M$ is also considered. 

\subsubsection{Point estimation of local extrema} 
We now turn to comparing the estimates $\hat{t}_i$ for $i = 1, \ldots, \hat{M}$ for each method. Since $\hat{M}$ might deviate from $M = 3$, as suggested in Figure~\ref{fig:freq}, we adopt the following convention to align the estimated local extrema with the true $t_i$ for $i = 1, 2, 3$ for all methods. We consider three intervals $(b_0, b_1)$, $(b_1, b_2)$, and $(b_2, b_3)$, where $b_0 = 0$, $b_3 = 1$, and $b_i = (t_i + t_{i + 1})/2$ for $i = 1, 2$. Then for each method, we collect all estimated local extrema that fall into each interval. If the $i$th interval ($i = 1, 2, 3$) contains more than one estimate, we use the average of all local extrema within the interval as the estimate of $t_i$; if an interval contains no estimates, we put an NA to indicate missingness. Performance of each method in estimating $t_i$ for $i = 1, 2, 3$ is compared by calculating the root mean squared error (RMSE), averaged across 100 simulations excluding NAs. The number of simulations in which a method gives zero or multiple local extrema within each interval is reported in Table~\ref{table:multiple}.

Table \ref{table:RMSE} reports the RMSE for estimated local extrema by all methods. The semiparametric Bayesian method DGP, with the Beta(1,1) or the Beta(2,3) prior, gives the smallest RMSEs in nearly all cases, with only one exception for $t_3$ at $n = 100$ when STS is slightly better. For $t_1$ and $t_2$, DGP often reduces the RMSEs of STS and NKS by over half or more, consistently across all sample sizes. For DGP, the two priors yield similar RMSEs in most cases, indicating that point estimates of local extrema tend to be minimally affected by the prior specification. Table~\ref{table:multiple} indicates that NKS produces multiple local extremum estimates in each interval much more often than DGP and STS, especially for $t_3$ with small curvature. When the sample size is 100, DGP leads to multiple local extrema in 14 (Beta(2,3) prior) and 15 (Beta(1, 1) prior) out of 100 simulations, while STS misses estimates within $(0, b_1)$ for 50 simulations. Both DGP and STS estimate local extrema that align well with the true local extrema when $n$ increases to 500 and 1000. It is worth mentioning that all methods give one local extremum in the interval $(b_1, b_2)$ for almost all simulations (Table~\ref{table:multiple}), which provides a scenario that eliminates the need to account for zero or multiple estimates; Table~\ref{table:RMSE} shows that in this scenario that corresponds to estimating $t_2$ the proposed DGP achieves the smallest RMSEs, suggesting superior performance of DGP. 

\begin{table}[ht]
\centering
\caption{Comparison of various methods using root mean square error (RMSE). The reported RMSEs are multiplied by 100 for easy comparison. The smallest and second smallest RMSEs in each column are marked in bold. All RMSEs are averaged across 100 repeated simulations.
} \label{table:RMSE}
\begin{tabular}{cccccccccc}
\toprule
\multirow{2}{*}{Method}  & \multicolumn{3}{c}{$n = 100$} & \multicolumn{3}{c}{$n = 500$} &  \multicolumn{3}{c}{$n = 1000$} \\
\cmidrule(lr){2-4} \cmidrule(lr){5-7} \cmidrule(lr){8-10} 
& $t_1$ & $t_2$ & $t_3$ & $t_1$ & $t_2$ & $t_3$ & $t_1$ & $t_2$ & $t_3$ \\
\cmidrule(lr){1-1} \cmidrule(lr){2-4} \cmidrule(lr){5-7} \cmidrule(lr){8-10}
DGP Beta(1,1) & \textbf{0.67} & \textbf{0.88} & 4.13 & \textbf{0.29} & \textbf{0.54} & \textbf{2.11} & \textbf{0.25} & \textbf{0.46} & \textbf{1.31} \\ 
DGP Beta(2,3) & \textbf{0.65} & \textbf{0.88} & \textbf{3.85} & \textbf{0.30} & \textbf{0.54} & \textbf{2.00} & \textbf{0.24} & \textbf{0.45} & \textbf{1.30} \\ 
STS & 1.65 & 1.53 & \textbf{3.14} & 0.78 & 1.22 & 2.56 & 0.75 & 1.11 & 1.90 \\ 
NKS & 1.68 & 1.46 & 4.98 & 1.54 & 1.08 & 3.30 & 1.90 & 1.07 & 2.25 \\ 
\bottomrule
\end{tabular}
\end{table}

\begin{table}[ht]
\centering
\caption{Number of simulations with missing or multiple estimated local extrema in each interval. The three intervals, indexed by $t_1, t_2,$ and $t_3$ in the table, are $(0, (t_1 + t_2)/2), ((t_1 + t_2)/2, (t_2 + t_3)/2), ((t_2 + t_3)/2, 1)$, respectively. The number of simulations with missingness, if non-zero, is reported as the second number in a pair; otherwise if there is no missingness, we only report the number of simulations with multiple estimates.} \label{table:multiple}
\begin{tabular}{cccccccccc}
\toprule
\multirow{2}{*}{Method}  & \multicolumn{3}{c}{$n = 100$} & \multicolumn{3}{c}{$n = 500$} &  \multicolumn{3}{c}{$n = 1000$} \\
\cmidrule(lr){2-4} \cmidrule(lr){5-7} \cmidrule(lr){8-10} 
& $t_1$ & $t_2$ & $t_3$ & $t_1$ & $t_2$ & $t_3$ & $t_1$ & $t_2$ & $t_3$ \\
\cmidrule(lr){1-1} \cmidrule(lr){2-4} \cmidrule(lr){5-7} \cmidrule(lr){8-10}
DGP Beta(1,1) & 0 & 0 & 15 & 0 & 0 & 5 & 0 & 0 & 1 \\ 
DGP Beta(2,3) & 0 & 0 & 14 & 0 & 0 & 1 & 0 & 0 & 0 \\ 
STS & (0, 50) & 0 & 1 & 1 & 0 & 1 & 1 & 0 & 0 \\ 
NKS & 19 & 1 & 47 & 14 & 3 & 45 & 20 & 0 & 40 \\
\bottomrule
\end{tabular}
\end{table}

We note that there are a few noticeable differences in our implementation of the encompassing strategy with respect to \cite{yu2023bayesian}. In our estimation approach, we fix the hyperparameters $\sigma, \tau$ and $h$ at the values that maximize the marginal unconstrained likelihood, while in the Monte Carlo Expectation Maximization (MCEM) approach of \cite{yu2023bayesian} $\sigma$ is sampled in each MC E-step while $\tau$ and $h$ are iteratively updated in the M-step given their previous values and the samples of $t$ and $\sigma$ drawn in the E-step. Furthermore, while we make use of the analytic form of the posterior distribution of $t$, the MCEM approach of \cite{yu2023bayesian} draws posterior samples of $t$. One clear advantage of our implementation is that it is computationally much faster than the MCEM method. (Using $n = 100$ as an example, our implementation was 100 times faster than MCEM on a regular PC, at a magnitude of 32 seconds versus 3200 seconds for completing 100 simulations.) When applying the MCEM method to the simulation study, we found overall similar values in the selected parameters. For example, estimates for $\sigma$, $\tau$ and $h$, averaged over 100 simulated datasets, were 0.10, 0.30 and 0.13, respectively, for our approach and 0.14, 0.37 and 0.13, respectively, for the MCEM.

\subsection{Interval estimation of local extrema}
We now assess the proposed interval estimates in~\eqref{eq:ci}. Within each HPDR segment under the Beta(1,1) prior, we estimate $\hat t_m$ by finding the local extremum of $\hat \mu_f$; if multiple local extrema are found, then the average is used. We assess the coverage of interval estimators conditional on $\hat{M} = 3$. This conditional event tends to occur with probability one given the consistency of $\hat{M}$ and indeed has a high probability in finite sample settings as observed in Section~\ref{sec:sim.M}. We consider three confidence levels $1 - \alpha$ for $\alpha \in \{0.1, 0.05, 0.1\}$. In addition to (marginal) confidence intervals for each $t_i$, we also obtain a joint confidence set for $\{t_1, t_2, t_3\}$ using the Bonferroni correction. We compare the empirical coverage of three marginal confidence intervals and one joint confidence set with the confidence level. 

Table~\ref{table:coverage} shows that the empirical coverage is close to the nominal level when $n$ increases to 500, for both marginal confidence intervals and joint confidence sets. We observe no significant derivation of the observed coverage from the confidence level relative to the standard errors when $n \in \{500, 1000\},$ indicating satisfactory coverage of the proposed interval estimates in this finite sample setting. In results not reported here, changing the prior to Beta(2,3) when deriving HPDR leads to a similar coverage for both marginal confidence intervals and joint confidence sets.

\begin{table}
\centering
\caption{Coverage of confidence intervals at various confidence levels $1 - \alpha$ for $\alpha \in \{0.1, 0.05, 0.01\}$. The first three blocks report the coverage of marginal confidence intervals for each local extremum, while the last block is the coverage of joint confidence sets using the Bonferroni correction. For each of the three rows, the maximum standard errors are 0.07, 0.04, and 0.04, respectively.} \label{table:coverage} 
\begin{tabular}{ccccccccccccc}
\toprule
& \multicolumn{3}{c}{$t_1$} & \multicolumn{3}{c}{$t_2$} & \multicolumn{3}{c}{$t_3$} & \multicolumn{3}{c}{Joint} \\ 
\cmidrule(lr){2-4} \cmidrule(lr){5-7} \cmidrule(lr){8-10} \cmidrule(lr){11-13}
& 0.1 & 0.05& 0.01 & 0.1 & 0.05& 0.01 & 0.1 & 0.05& 0.01 & 0.1 & 0.05& 0.01  \\ 
\cmidrule(lr){2-4} \cmidrule(lr){5-7} \cmidrule(lr){8-10} \cmidrule(lr){11-13} 
$n=100$ 	&0.81	&0.91	&0.98	&0.75 	&0.81	&0.87	 &0.74	&0.81	&0.83	&0.66	&0.72	&0.77 \\
$n=500$ 	&0.91	&0.94	&1	&0.84	&0.90	&0.99	&0.90	&0.94	&1	&0.85	&0.94	&1 \\
$n=1000$ 	&0.86	&0.95	&0.99	&0.86	&0.92	&0.99	&0.88	&0.94	&0.99	&0.85	&0.93	&0.99 \\
\bottomrule
\end{tabular}
\end{table}

\section{Real data application} \label{sec:application} 

In this section, we show an application of our method to the analysis of event-related potentials (ERP), which represent electroencephalogram (EEG) recorded in response to stimuli. Primary statistical analyses of an ERP waveform focus on estimating the amplitude (microvolts) and latency (milliseconds) of specific peaks and dips, also called ERP components, as these have been shown to be associated with human sensory and cognitive functions \citep{Luck2005}. Although ERPs have been extensively used in psychology and the cognitive science community, research in statistical modeling for latency estimation with uncertainty quantification is not mature yet and still under development. Here we show how our methodology can be applied to derive posterior distributions of ERP component latencies, an important information when making scientific discoveries based on ERP data \citep{yu2023bayesian}.

\begin{figure}
\centering
\includegraphics[width=5in]{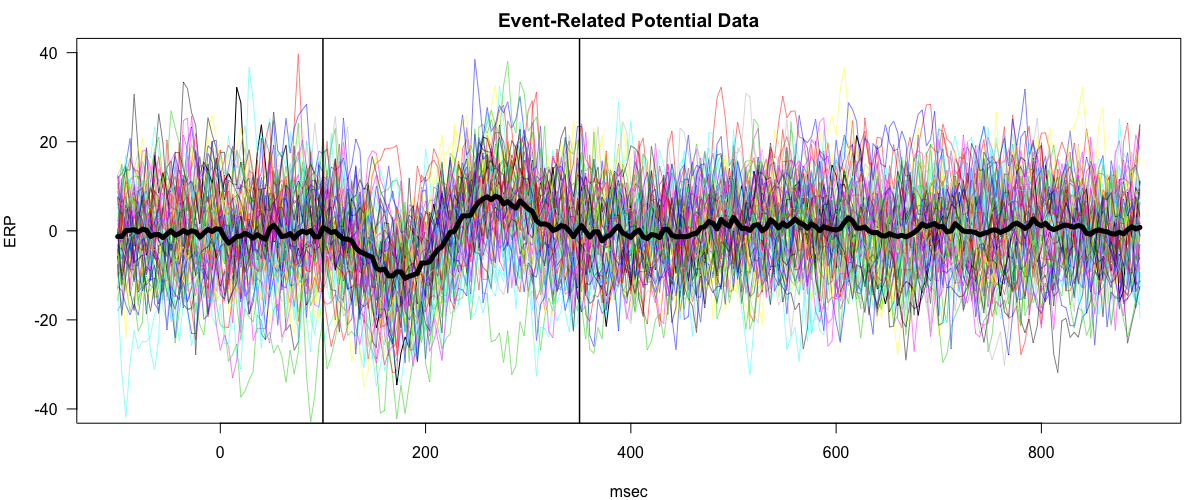}
\caption{ERP data: 72 time series each corresponding to one single trial and the grand average time course averaged over all 72 trials. The time epoch between the two vertical lines defines the search window for components N1 and P3.}
\label{fig:erp_data}
\end{figure} 

We use ERP data publicly available at \url{http://dsenturk.bol.ucla.edu/supplements.html}. The dataset consists of ERP signals of a single subject with autism spectrum disorder (ASD), evaluated at one electrode, one condition, and 72 trials, each having 250 time points. Figure \ref{fig:erp_data} shows the time series of all 72 trials and the grand average time course averaged over all 72 trials. Two ERP components, N1, typically within the window [100, 250] msec, and P3, typically within the window [190, 350] msec, are the main interest of the study.  We therefore restrict our analysis to the time window [100, 350] msec.

\begin{figure}
\centering
\includegraphics[width=4in]{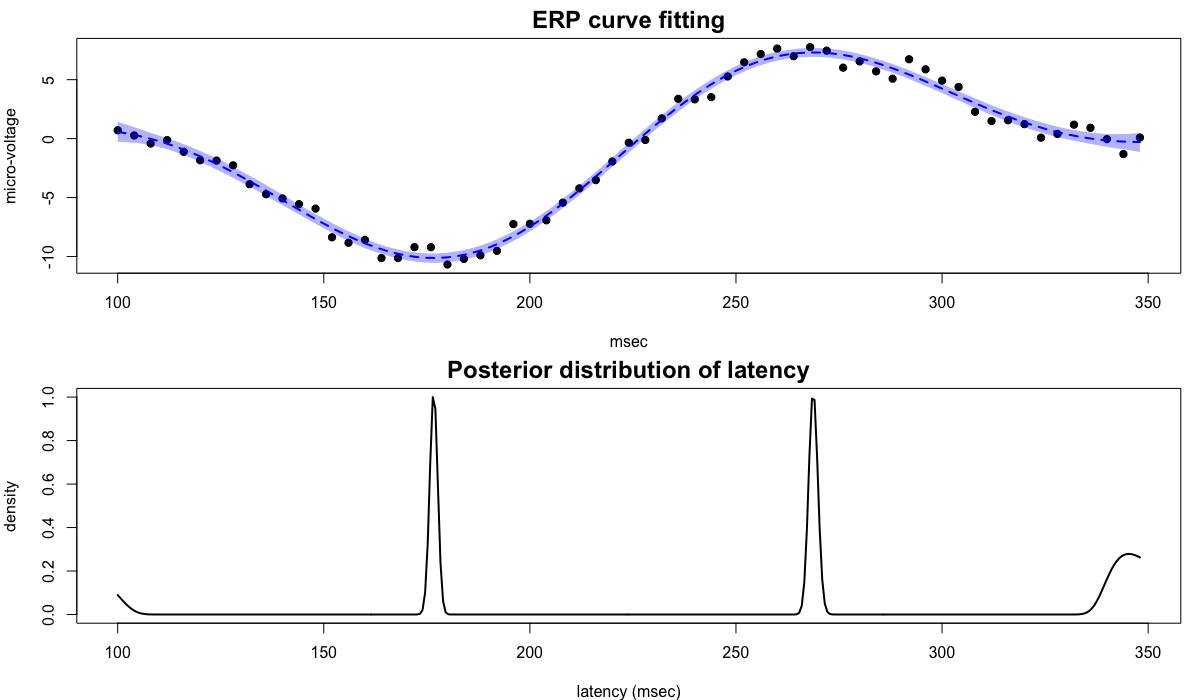}
\caption{ERP data: Curve fitting and posterior density of the latency of the grand ERP waveform averaged over 72 trials.}
\label{fig:erp_fitting_post_latency}
\end{figure}

Since EEG signals are typically noisy, traditionally neuroscientists average signals across trials to obtain an overall or grand average ERP waveform, which they then visually inspect to determine the amplitude size and latency location of the ERP components. Following such practice, we first applied our method to the grand average time course. As in the simulations, empirical Bayes estimates of the parameters other than $t$ were obtained by maximizing the marginal likelihood, as $(\sigma, \tau, h) = (0.642, 6.385, 0.053)$, with a uniform prior on $t$. Curve fitting and posterior distribution of the latency are shown in Figure \ref{fig:erp_fitting_post_latency}. We note that the observations, i.e. the grand average over all trials, and the fitted curve appear to have a similar smoothness level as in the preceding simulation section; for example, compare Figure~\ref{fig:erp_fitting_post_latency} (top row) with Figure~\ref{fig:simulated_data_100}. In addition to generating a smooth fitted ERP curve along with 95\% credible intervals for amplitude estimation, our model-based approach provides a full posterior distribution of latency locations for ERP components. The 95\% credible intervals for the N1 and P3 latencies are [174.58, 178.32] msec and [266.58, 270.93] msec, respectively.

\begin{figure}
\centering
\includegraphics[width=3in]{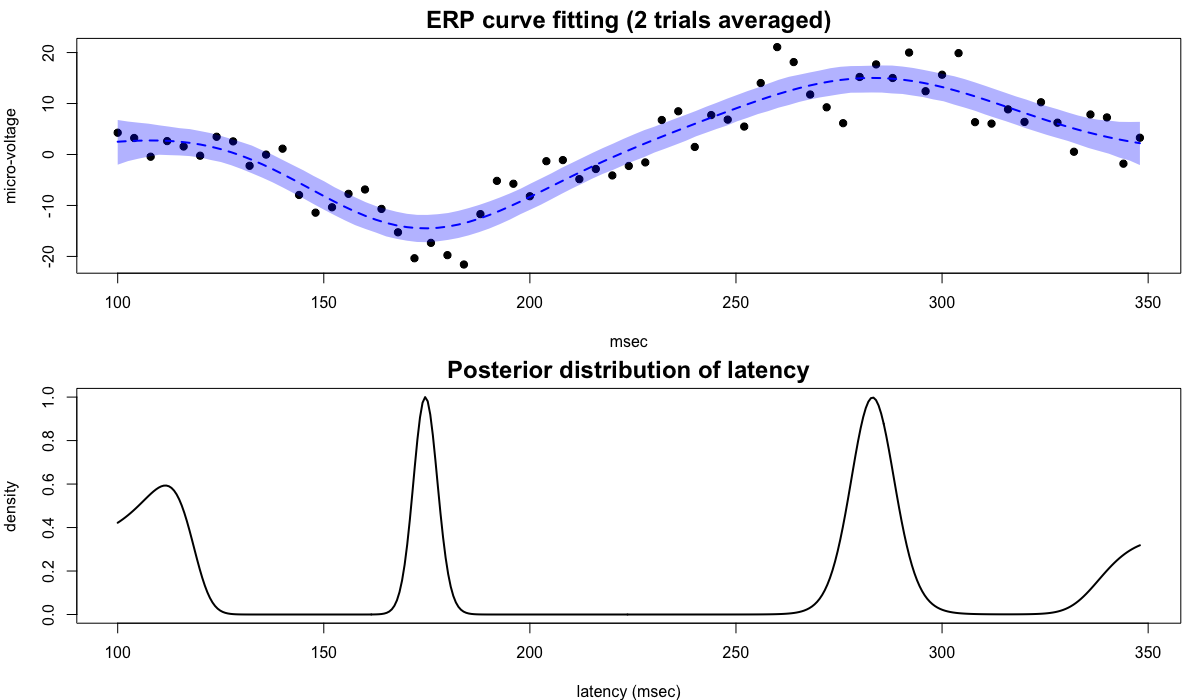}
\includegraphics[width=3in]{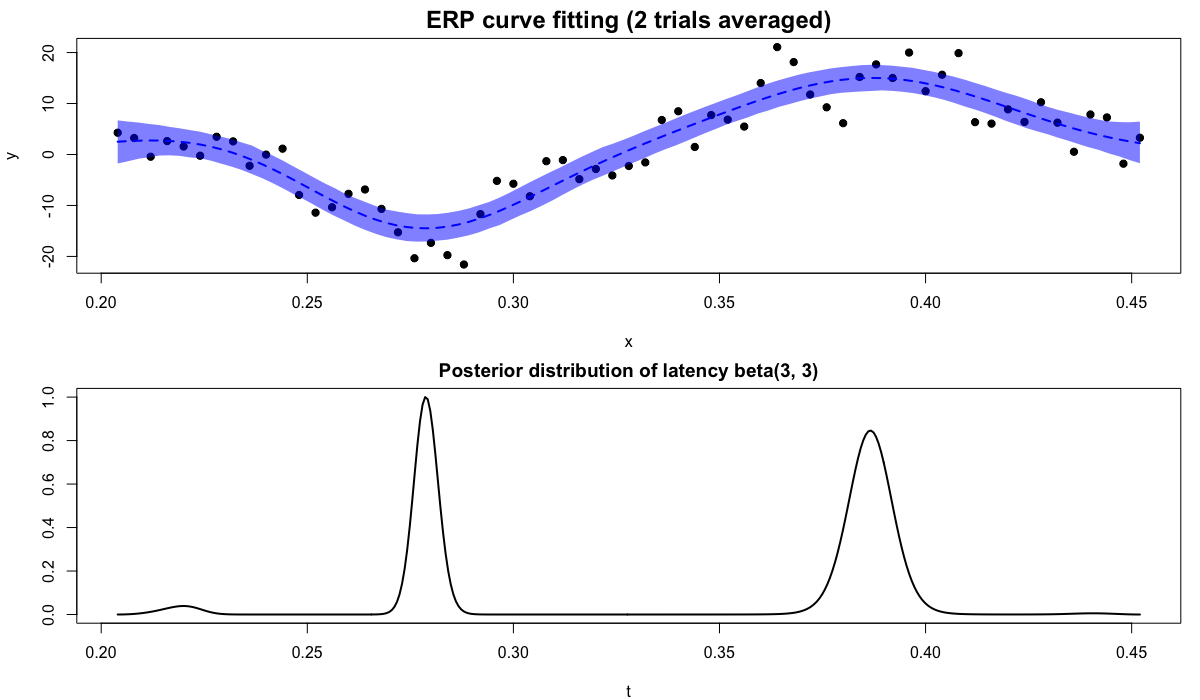}
\caption{ERP data: Curve fitting and posterior density of the latency of the ERP waveform averaged over the first two trials, with a uniform prior (left panel) and a Beta(3,3) prior (right panel) on the latency.}
\label{fig:erp_fitting_post_latency_two_trial}
\end{figure}

We also investigated the robustness of the results to the smoothness of the ERP waveform. Indeed, since our model explicitly accounts for errors in the data, some of the
excessive averaging, which is routinely done to obtain smooth curves, can be avoided. When fewer trials are averaged, we expect posterior distributions with larger variation. As an example, the left panel of Figure \ref{fig:erp_fitting_post_latency_two_trial} shows the estimated waveform and the  posterior density of latency when only the first 2 trials are averaged. The N1 and P3 latencies are still identified, though, as expected, with larger uncertainty. In particular, the 95\% credible intervals for N1 and P3 are [168.99, 180.80] msec and [271.55, 295.17] msec, respectively. Furthermore, as an effect of the smaller level of averaging, some smaller modes at the extremes of the interval are now more pronounced. These are spurious effects and can be avoided by utilizing a prior distribution on the latency that discourages local extrema at the endpoints of the interval. For example, the right panel of Figure  \ref{fig:erp_fitting_post_latency_two_trial} shows the inference using a Beta(3, 3) prior. 

\section{Discussion} \label{sec:conclusion}
In this article, we have studied an encompasssing semiparametric Bayesian approach for identifying multiple local extrema of an unknown function. We have shown that the posterior distribution is connected to unconstrained nonparametric regression in a closed-form characterization, enabling fast computation. We have established local asymptotic normality properties and convergence to Gaussian mixtures for this unconventional strategy, indicating multi-modality of the posterior distribution and substantiating the use of the highest posterior density region for posterior exploration. Our simulations have suggested superior performance of this encompassing semiparametric method relative to existing methods. 

Although we have focused on Gaussian processes with stationary covariance functions whose eigenvalues decay at certain rates as special examples, the developed framework of this article, which bases inference on the multi-modal posterior distribution of $t$ with justified asymptotic properties, can be extended to other nonparametric priors, including Gaussian processes with other covariance functions and random series priors. Similar to the flexibility encoded in covariance kernels of Gaussian processes, the rich menu for basis functions in random series priors allows flexible shapes of the underlying functions; for example, wavelets might be better suited for spiky functions in certain applications such as mass spectrometry \citep{Liu2020fqr}, and B-splines for locally supported functions~\citep{wang2023functional}. In these generalizations, one needs to verify the conditions in Theorem~\ref{thm:bvm} for the adopted nonparametric prior, with technical challenges including deriving the approximate properties of relevant estimates as in Lemma~\ref{thm:nonasy.bound} and Assumption C, and selecting hyperparameters such as the number of basis functions in the context of local extrema detection.

Throughout the paper we have focused on a one-dimensional sample space. It may be argued that the encompassing strategy studied in this article generalizes to $d$-dimensional compact sample spaces for any $d \geq 1$ by using GP prior counterparts supported on $d$-dimensional $\mX$. However, the main challenges in multivariate settings include the need to theoretically study multivariate posterior distributions with multi-modality, and develop computationally efficient algorithms for posterior exploration.

There are several other interesting future directions to pursue. Firstly, Assumption A3 can be relaxed to allow local extrema based on high-order derivative tests, and we envision the developed arguments in this article are largely applicable with the LAN expansion extended to its higher-order counterpart. Secondly, one may study the encompassing strategy with potentially different posterior exploration methods when there are many or even a diverging number of local extrema, and compare its performance with alternative approaches. Finally, one substantial challenge the proposed method overcomes is the multiplicity of local extrema with unknown dimensions and locations. With given $M$, which is a different setting, further efficiency gain might be possible by incorporating this knowledge into the method. In this case, it is also interesting to study the optimal rate for estimating the $M$-dimensional local extrema, and compare the proposed estimator with the optimal rate.

\normalem
\bibliographystyle{apalike}
\bibliography{reference}

\begin{thebibliography}{}

\bibitem[Abraham and Khadraoui, 2015]{Abraham2015}
Abraham, C. and Khadraoui, K. (2015).
\newblock Bayesian regression with {B}-splines under combinations of shape
  constraints and smoothness properties.
\newblock {\em Statistica Neerlandica}, 69:150--170.

\bibitem[Carreira-Perpin{\'a}n and Williams, 2003]{carreira2003number}
Carreira-Perpin{\'a}n, M.~A. and Williams, C.~K. (2003).
\newblock On the number of modes of a {G}aussian mixture.
\newblock In {\em International Conference on Scale-Space Theories in Computer
  Vision}, pages 625--640. Springer.

\bibitem[Castillo, 2012]{castillo2012semiparametric}
Castillo, I. (2012).
\newblock A semiparametric {B}ernstein--von {M}ises theorem for {G}aussian
  process priors.
\newblock {\em Probability Theory and Related Fields}, 152(1-2):53--99.

\bibitem[Castillo and Nickl, 2014]{castillo2014bernstein}
Castillo, I. and Nickl, R. (2014).
\newblock On the {B}ernstein--von {M}ises phenomenon for nonparametric {B}ayes
  procedures.
\newblock {\em The Annals of Statistics}, 42(5):1941--1969.

\bibitem[Castillo and Rousseau, 2015]{castillo2015bernstein}
Castillo, I. and Rousseau, J. (2015).
\newblock A {B}ernstein--von {M}ises theorem for smooth functionals in
  semiparametric models.
\newblock {\em The Annals of Statistics}, 43(6):2353--2383.

\bibitem[Castillo et~al., 2015]{Castillo2015linearmodel}
Castillo, I., Schmidt-Hieber, J., and van~der Vaart, A.~W. (2015).
\newblock Bayesian linear regression with sparse priors.
\newblock {\em The Annals of Statistics}, 43:1986--2018.

\bibitem[Cheng and Schwartzman, 2017]{Cheng2017}
Cheng, D. and Schwartzman, A. (2017).
\newblock {Multiple testing of local maxima for detection of peaks in random
  fields}.
\newblock {\em The Annals of Statistics}, 45(2):529--556.

\bibitem[Conway, 1994]{conway1994course}
Conway, J. (1994).
\newblock {\em A Course in Functional Analysis}.
\newblock Graduate Texts in Mathematics. Springer New York.

\bibitem[Cucker and Zhou, 2007]{cucker2007learning}
Cucker, F. and Zhou, D.-X. (2007).
\newblock {\em Learning Theory: An Approximation Theory Viewpoint}, volume~24.
\newblock Cambridge University Press.

\bibitem[Dasgupta et~al., 2021]{dasgupta2021modality}
Dasgupta, S., Pati, D., Jermyn, I.~H., and Srivastava, A. (2021).
\newblock Modality-constrained density estimation via deformable templates.
\newblock {\em Technometrics}, 63(4):536--547.

\bibitem[Davies et~al., 2001]{davies2001local}
Davies, P., Kovac, A., et~al. (2001).
\newblock Local extremes, runs, strings and multiresolution.
\newblock {\em The Annals of Statistics}, 29(1):1--65.

\bibitem[Devroye et~al., 2018]{devroye2018total}
Devroye, L., Mehrabian, A., and Reddad, T. (2018).
\newblock The total variation distance between high-dimensional {G}aussians
  with the same mean.
\newblock {\em arXiv preprint arXiv:1810.08693}.

\bibitem[Egner et~al., 2007]{egner2007fluorescence}
Egner, A., Geisler, C., Von~Middendorff, C., Bock, H., Wenzel, D., Medda, R.,
  Andresen, M., Stiel, A.~C., Jakobs, S., Eggeling, C., Sch\"onle, A., and
  Hell, S.~W. (2007).
\newblock Fluorescence nanoscopy in whole cells by asynchronous localization of
  photoswitching emitters.
\newblock {\em Biophysical journal}, 93(9):3285--3290.

\bibitem[Ferreira and Menegatto, 2012]{Ferreira2012}
Ferreira, J.~C. and Menegatto, V.~A. (2012).
\newblock {Reproducing properties of differentiable Mercer-like kernels}.
\newblock {\em Mathematische Nachrichten}, 285(8-9):959--973.

\bibitem[Geisler et~al., 2007]{geisler2007resolution}
Geisler, C., Sch{\"o}nle, A., Von~Middendorff, C., Bock, H., Eggeling, C.,
  Egner, A., and Hell, S. (2007).
\newblock Resolution of $\lambda$/10 in fluorescence microscopy using fast
  single molecule photo-switching.
\newblock {\em Applied Physics A}, 88(2):223--226.

\bibitem[Ghosal and van~der Vaart, 2017]{Ghosal+van:17}
Ghosal, S. and van~der Vaart, A. (2017).
\newblock {\em {Fundamentals of Nonparametric Bayesian Inference}}.
\newblock Cambridge University Press, Cambridge.

\bibitem[Gin{\'e} and Nickl, 2021]{gine2021mathematical}
Gin{\'e}, E. and Nickl, R. (2021).
\newblock {\em Mathematical Foundations of Infinite-dimensional Statistical
  Models}.
\newblock Cambridge university press.

\bibitem[Holmes and Heard, 2003]{Holmes2003}
Holmes, C.~C. and Heard, N.~A. (2003).
\newblock Generalized monotonic regression using random change points.
\newblock {\em Statistics in Medicine}, 22:623--638.

\bibitem[Kim, 2006]{kim2006bernstein}
Kim, Y. (2006).
\newblock The {B}ernstein--von {M}ises theorem for the proportional hazard
  model.
\newblock {\em The Annals of Statistics}, 34(4):1678--1700.

\bibitem[Kim and Lee, 2004]{kim2004bernstein}
Kim, Y. and Lee, J. (2004).
\newblock A {B}ernstein-von {M}ises theorem in the nonparametric
  right-censoring model.
\newblock {\em The Annals of Statistics}, 32(4):1492--1512.

\bibitem[Kleijn and van~der Vaart, 2012]{kleijn2012bernstein}
Kleijn, B. J.~K. and van~der Vaart, A.~W. (2012).
\newblock The {B}ernstein-von-{M}ises theorem under misspecification.
\newblock {\em Electronic Journal of Statistics}, 6:354--381.

\bibitem[Kovac, 2007]{Kovac2007}
Kovac, A. (2007).
\newblock {Smooth functions and local extreme values}.
\newblock {\em Computational Statistics and Data Analysis}, 51(10):5155--5171.

\bibitem[Liu et~al., 2020]{Liu2020fqr}
Liu, Y., Li, M., and Morris, J.~S. (2020).
\newblock {Function-on-scalar quantile regression with application to mass
  spectrometry proteomics data}.
\newblock {\em Annals of Applied Statistics}, 14(2):521--541.

\bibitem[Liu and Li, 2022]{liu2022optimal}
Liu, Z. and Li, M. (2022).
\newblock Optimal plug-in {G}aussian processes for modelling derivatives.
\newblock {\em arXiv preprint arXiv:2210.11626}.

\bibitem[Liu and Li, 2023]{liu2023estimation}
Liu, Z. and Li, M. (2023).
\newblock On the estimation of derivatives using plug-in kernel ridge
  regression estimators.
\newblock {\em Journal of Machine Learning Research}.
\newblock In press.

\bibitem[Luck, 2005]{Luck2005}
Luck, S.~J. (2005).
\newblock {\em An Introduction to the Event-Related Potential Technique}.
\newblock The MIT Press.

\bibitem[Meyer, 2008]{Meyer2008}
Meyer, M.~C. (2008).
\newblock Inference using shape-restricted regression splines.
\newblock {\em Annals of Applied Statistics}, 2:1013--1033.

\bibitem[Neelon and Dunson, 2004]{Neelon2004}
Neelon, B. and Dunson, D.~B. (2004).
\newblock Bayesian isotonic regression and trend analysis.
\newblock {\em Biometrics}, 60:398--406.

\bibitem[Pati and Bhattacharya, 2015]{pati2015adaptive}
Pati, D. and Bhattacharya, A. (2015).
\newblock Adaptive {B}ayesian inference in the {G}aussian sequence model using
  exponential-variance priors.
\newblock {\em Statistics \& Probability Letters}, 103:100--104.

\bibitem[Raghuraman et~al., 2001]{raghuraman2001replication}
Raghuraman, M., Winzeler, E.~A., Collingwood, D., Hunt, S., Wodicka, L.,
  Conway, A., Lockhart, D.~J., Davis, R.~W., Brewer, B.~J., and Fangman, W.~L.
  (2001).
\newblock Replication dynamics of the yeast genome.
\newblock {\em Science}, 294(5540):115--121.

\bibitem[Ramsay, 1998]{Ramsay1998}
Ramsay, J.~O. (1998).
\newblock Estimating smooth monotone functions.
\newblock {\em Journal of the Royal Statistical Society: Series B (Statistical
  Methodology)}, 60(2):365--375.

\bibitem[Rasmussen and Williams, 2006]{rasmussen2006}
Rasmussen, C.~E. and Williams, C.~K. (2006).
\newblock {\em Gaussian Process for Machine Learning}.
\newblock The MIT Press.

\bibitem[Schwartzman et~al., 2011]{Schwartzman2011}
Schwartzman, A., Gavrilov, Y., and Adler, R.~J. (2011).
\newblock {Multiple testing of local maxima for detection of peaks in 1D}.
\newblock {\em The Annals of Statistics}, 39(6):3290--3319.

\bibitem[Shively et~al., 2009]{Shively2009}
Shively, T.~S., Sager, T.~W., and Walker, S.~G. (2009).
\newblock A {B}ayesian approach to non-parametric monotone function estimation.
\newblock {\em Journal of the Royal Statistical Society: Series B (Statistical
  Methodology)}, 71(1):159--175.

\bibitem[Shively et~al., 2011]{Shively2011}
Shively, T.~S., Walker, S.~G., and Damien, P. (2011).
\newblock Nonparametric function estimation subject to monotonicity, convexity
  and other shape constraints.
\newblock {\em Journal of Econometrics}, 161:166--181.

\bibitem[Song et~al., 2006]{Song2006}
Song, P., Gao, X., Liu, R., and Le, W. (2006).
\newblock Nonparametric inference for local extrema with application to
  oligonucleotide microarray data in yeast genome.
\newblock {\em Biometrics}, 62(2):545--554.

\bibitem[van~der Vaart, 2000]{van2000asymptotic}
van~der Vaart, A.~W. (2000).
\newblock {\em Asymptotic statistics}, volume~3.
\newblock Cambridge university press.

\bibitem[Wahba, 1990]{wahba1990spline}
Wahba, G. (1990).
\newblock {\em Spline Models for Observational Data}.
\newblock SIAM.

\bibitem[Wang et~al., 2023]{wang2023functional}
Wang, Z., Magnotti, J., Beauchamp, M.~S., and Li, M. (2023).
\newblock Functional group bridge for simultaneous regression and support
  estimation.
\newblock {\em Biometrics}, 79(2):1226--1238.

\bibitem[Wheeler et~al., 2017]{wheeler2017bayesian}
Wheeler, M.~W., Dunson, D.~B., and Herring, A.~H. (2017).
\newblock Bayesian local extremum splines.
\newblock {\em Biometrika}, 104(4):939--952.

\bibitem[Yoo and Ghosal, 2016]{yoo2016supremum}
Yoo, W.~W. and Ghosal, S. (2016).
\newblock Supremum norm posterior contraction and credible sets for
  nonparametric multivariate regression.
\newblock {\em The Annals of Statistics}, 44(3):1069--1102.

\bibitem[Yu et~al., 2023]{yu2023bayesian}
Yu, C.-H., Li, M., Noe, C., Fischer-Baum, S., and Vannucci, M. (2023).
\newblock Bayesian inference for stationary points in {G}aussian process
  regression models for event-related potentials analysis.
\newblock {\em Biometrics}, 79(2):629--641.

\end{thebibliography}

\newpage 
\renewcommand{\thepage}{S\arabic{page}}  
\setcounter{page}{1}
\appendix
\begin{center}
\Large
Supplementary material for ``Semiparametric Bayesian inference for local extrema of functions in the presence of noise"
\end{center}

In this supplementary material, we present proofs of all results in the main paper, additional technical lemmas, and additional numerical experiments. 

\section{Proofs} \label{sec:main.proof}

\subsection{Proof of Proposition~\ref{lem:closed-form}}

By Bayes' theorem, it suffices to show that the likelihood takes the form of \eqref{eq:likelihood}. 
Recall that $\bm{y} | X,t \sim N(0, \Sigma_t)$, where $\Sigma_t=\sigma^2 (n \lambda)^{-1}(A+B)$ with $A = K(X,X) + n \lambda \bI_n$ and $B = -  K_{01}(X, t)K^{-1}_{11}(t, t) K_{10}(t, X) = -\bm{a} \bm{a}^T$ by letting $\bm{a} = K_{01}(X, t) K_{11}^{-1/2}(t, t) $. Note that the condition $\widehat{\sigma}^2_{f'}(t) > 0$ for any $t$ ensures $K_{11}(t, t) > 0$ in view of~\eqref{eq:post.var.derivative}.

In view of the Sherman–Morrison formula, we have $\det{(A + B)} = (1 - \bm{a}^T A^{-1} \bm{a}) \det(A)$ and $(A + B)^{-1} = A^{-1} + \frac{A^{-1} \bm{a} \bm{a}^T A^{-1}}{1 - \bm{a}^T A^{-1} \bm{a}}$, assuming $1 - \bm{a}^T A^{-1} \bm{a} \neq 0$. Substituting these two identities into the multivariate normal density $\ell(t)$ yields 
\begin{align}
\ell(t) & = \{2 \pi \sigma^2(n \lambda)^{-1}\}^{-n/2} \det(A + B)^{-1/2} \exp\left \{ -\frac{1}{2 \sigma^2(n \lambda)^{-1}} \by^T (A + B)^{-1} \by \right \} \\
& = \{2 \pi \sigma^2(n \lambda)^{-1}\}^{-n/2} (\det(A))^{-1/2} \exp\left\{-\frac{\by^T A^{-1} \by}{2\sigma^2(n \lambda)^{-1}} \right\} \\
& \quad\ \cdot (1 - \bm{a}^T A^{-1} \bm{a})^{-1/2} \exp\left\{  -\frac{\by^T A^{-1} \bm{a} \bm{a}^T A^{-1} \by }{2 \sigma^2(n \lambda)^{-1} (1 - \bm{a}^T A^{-1} \bm{a})} \right\} \\
& =  C \{\sigma^2 (n \lambda)^{-1} (1 - \bm{a}^T A^{-1} \bm{a})\}^{-1/2} \exp\left \{  -\frac{\by^T A^{-1} \bm{a} K_{11}(t, t)\bm{a}^T A^{-1} \by}{2 \sigma^2 (n \lambda)^{-1} K_{11}(t, t) (1 - \bm{a}^T A^{-1} \bm{a})} \right \},
\end{align} 
where $$ C = \{2 \pi \sigma^2(n \lambda)^{-1}\}^{-n/2} (\det(A))^{-1/2} \exp\left\{-\frac{\by^T A^{-1} \by}{2\sigma^2(n \lambda)^{-1}} \right\} \cdot \{\sigma^2 (n \lambda)^{-1} \}^{1/2}$$
does not depend on $t$. The proof is completed by noticing that $\widehat{\mu}_{f'}(t) = K_{11}^{1/2}(t, t) \bm{a}^T A^{-1} \by$ and $\widehat{\sigma}^2_{f'}(t) = \sigma^2 (n \lambda)^{-1} K_{11}(t, t) (1 - \bm{a}^TA^{-1} \bm{a})$. This completes the proof.

\subsection{Proof of Lemma~\ref{thm:nonasy.bound}}
A one-dimensional version of Theorem 4 in \cite{liu2023estimation} shows that
\begin{align}\label{eq:lemma2.1}
\|\widehat{\mu}^{(k)}_{f'} - f^{(k+1)}_{\lambda}\|_{\infty} \leq&\frac{\sqrt{\kappa \kappa_{k+1,k+1}} \|f_0\|_\infty \sqrt{\log(9/\delta)}}{\sqrt{n}\lambda}  \left(10 + \frac{4 \kappa\sqrt{\log(9/\delta)}}{3 \sqrt{n\lambda}} \right)\\
&+\frac{C_2 \sqrt{\kappa \kappa_{k+1,k+1}} \sigma \sqrt{\log(3/\delta)}}{\sqrt{n} \lambda},\quad 0\leq k\leq 3,
\end{align}
For any bounded $f \in \Ltwo$, we define a bias of estimators of $f$ by matrix and integral operation as
\begin{align}\label{eq:def.e}
E(K, X, f) & = (L_{K, X} + \lambda I)^{-1} L_{K, X} f - (L_K + \lambda I)^{-1} L_K  f  \\
\label{eq:def.E} &= K(\cdot, X)[K(X, X) + n \lambda \bm{I}_n]^{-1}f(X) - (L_K + \lambda I)^{-1} L_K  f, 
\end{align}
which belongs to $\bbH$. Consider any $j,l\geq 1$ and $j+l\leq 5$, taking $f = K_{0l, x}$ yields
\begin{equation}
\partial^j E(K, X, K_{0l, x}) =  K_{j0}(\cdot, X)[K(X, X) + n \lambda \bI_n]^{-1} K_{0l}(X, x) - \partial^j (L_K + \lambda I)^{-1}L_K K_{0l, x}.
\end{equation}
Thus,
\begin{align}
\partial^j E(K, X, K_{0l, x})(x) & =  K_{j0}(x, X)[K(X, X) + n \lambda \bI_n]^{-1} K_{0l}(X, x) - \partial^j (L_K + \lambda I)^{-1}L_K K_{0l, x}(x) \\
& = K_{j0}(x, X)[K(X, X) + n \lambda \bI_n]^{-1} K_{0l}(X, x)  - (L_K + \lambda I)^{-1}L_K K_{jl, x}(x)
\end{align}
We write $(L_K + \lambda I)^{-1}L_K K_{jl,x}(x)=K_{jl,x}(x)-\la(L_K + \lambda I)^{-1} K_{jl,x}(x)=K_{jl}(x,x)-\la \varphi_{jl}(x)$. Then, by Theorem 16 in \cite{liu2023estimation} we have that for any $\delta\in(0,1)$, with $\PP_0$-probability at least $1-\delta$ it holds
\begin{align}
&\ |K_{j0}(x, X)[K(X, X) + n \lambda \bI_n]^{-1} K_{0l}(X, x)  - K_{jl}(x,x)+\la \varphi_{jl}(x) | \\
\leq &\ \|\partial^j E(K, X, K_{0l, x}) \|_{\infty}\\
\leq &\ \frac{ \sqrt{\kappa \kappa_{jj}} \|K_{0l, x}\|_\infty \sqrt{\log(3/\delta)}}{\sqrt{n}\lambda}  \left(10 + \frac{4 \sqrt{\kappa}\sqrt{\log(3/\delta)}}{3 \sqrt{n\lambda}} \right)\\
=&\ \frac{\sqrt{\kappa \kappa_{jj}} \kappa_{0l} \sqrt{\log(3/\delta)}}{\sqrt{n}\lambda}  \left(10 + \frac{4 \sqrt{\kappa}\sqrt{\log(3/\delta)}}{3 \sqrt{n\lambda}} \right).
\end{align}
In view of \eqref{eq:sigma.higher.deriv}, $\widehat{\sigma}^{2(k)}_{f'}(x)$ is a linear combination of quadratic forms
\begin{equation}
K_{jl}(x,x)-K_{j0}(\cdot, X)[K(X, X) + n \lambda \bI_n]^{-1} K_{0l}(X, x).
\end{equation}
Therefore, for any $0\leq k\leq 3$, we have
\begin{align}\label{eq:lemma2.3}
&|\widehat{\sigma}^{2(k)}_{f'}(x)-\sigma^2n^{-1}\sum_{i=0}^{k}{k\choose i}\varphi_{i+1,k+1-i}(x)| \\
&\qquad\qquad \leq \sum_{i=0}^{k}{k\choose i}\left[\frac{\sqrt{\kappa \kappa_{i+1,i+1}} \kappa_{0,k+1-i}\sigma^2 \sqrt{\log(3/\delta)}}{n\sqrt{n}\lambda^2}  \left(10 + \frac{4 \sqrt{\kappa}\sqrt{\log(3/\delta)}}{3 \sqrt{n\lambda}}\right)\right].
\end{align}
The above \eqref{eq:lemma2.1} and \eqref{eq:lemma2.3} can hold simultaneously with $\PP_0$-probability $1-8\delta$. Let $8\delta=n^{-10}$, and $A_n$ be the corresponding event. We immediately have $\PP_0(A_n)\geq 1-n^{-10}$ with $\log(3/\delta) \leq 10\log n + 4$ and $\log(9/\delta) \leq 10\log n + 5$ in the upper bound. This completes the proof.

\subsection{Proof of Lemma~\ref{lem:post.mode}}
First we prove that for any local extremum $t_m$ of $f_0$, there exists a local extremum $t_{\la,m}$ of $f_\la$ such that $t_{\la,m}\rightarrow t_m$ as $\la\rightarrow 0$. There exists $\delta>0$ such that for any $0<\epsilon<\delta$, it holds that $f'_0(t_m-\epsilon)<0$, $f'_0(t_m+\epsilon)>0$ and $f''_0(t_m\pm\epsilon)\neq 0$ without loss of generality. By Assumption C, we have
\begin{equation}
|f'_\la(t_m-\epsilon)-f'_0(t_m-\epsilon)|\lesssim\la^{r_1}.
\end{equation}	
Hence, for sufficiently small $\lambda$, it holds $f'_\la(t_m-\delta/2)<0$. Similarly, we have $f'_\la(t_m+\delta/2)>0$. According to the continuity of $f_\la$, there exists a $t_{\la,m}\in(t_m-\delta/2, t_m+\delta/2)$ such that $f'_\la(t_{\la,m})=0$. It can also be shown that $f_\la''(t)\neq 0$ for any $t\in(t_m-\delta/2,t_m+\delta/2)$ and sufficiently small $\la$, which implies $f_\la''(t_{\la,m})\neq 0$. Finally, we have $t_{\la,m}\rightarrow t_m$ as $\delta\rightarrow0$ and $\la\rightarrow 0$.

Again by Assumption C we can see that
\begin{equation}
|f'_\la(t_{\la,m})-f'_0(t_{\la,m})|\lesssim\la^{r_1}.
\end{equation}
Since $f'_\la(t_{\la,m})=0$, in view of the mean value theorem, we have
\begin{equation}
|f'_0(t_m)+f''_0(\xi_1)(t_{\la,m}-t_m)|\lesssim \la^{r_1},
\end{equation}
where $\xi_1$ lies between and $t_{\la,m}$ and $t_m$. Since $\xi_1\rightarrow t_m$ and $f_0''(t_m)\neq 0$ by Assumption A, we obtain
\begin{equation}
|t_{\la,m}-t_m|\lesssim \la^{r_1}.
\end{equation}

Under $A_n$, the existence and convergence rate of $\hat t_m$ can be shown similarly by applying \eqref{eq:dev.sigma.hat.sigma}. This completes the proof.

\subsection{Proof of Theorem~\ref{thm:LAN}}
The proof is based on the high probability event $A_n$ defined in Lemma~\ref{thm:nonasy.bound}. Conditions of Theorem~\ref{thm:LAN} imply $n^{\frac 12-2\beta}\varphi_{11,m}=o(1)$, yielding $\widehat{\sigma}^2_{f'}(t_{\la,m})>0$ in view of \eqref{eq:sigma.t.hat}. Invoking the likelihood function $\ell(t)$ in \eqref{eq:likelihood}, which holds at $t_{\la, m}$ and in its small neighborhood, we have
\begin{align}
\Lambda(t, \Delta t) = \log \frac{\ell(t + \Delta t)}{\ell(t)}= \frac{\ell'(t)}{\ell(t)}\Delta t+\frac{\ell''(t)\ell(t)-\ell'(t)^2}{2\ell(t)^2}(\Delta t)^2+R_3(\xi)(\Delta t)^3,
\end{align}
where $R_3(\xi)=\{2\ell'(t)^3+\ell'''(\xi)\ell(\xi)^2-3\ell'(\xi)\ell''(\xi)\ell(\xi)\}/\{6\ell(\xi)^3\}$ and $\xi$ is bewteen $t$ and $t+\Delta t$. Thus,
\begin{align}
\Lambda(t_{\la,m},\frac{u}{n^\beta})&=\left[-\frac{\widehat{\sigma}^{2'}_{f'}(t_{\la,m})}{2\widehat{\sigma}^2_{f'}(t_{\la,m})}-\frac{\widehat{\mu}_{f'}(t_{\la,m})\widehat{\mu}_{f'}'(t_{\la,m})}{\widehat{\sigma}^2_{f'}(t_{\la,m})}+\frac{\widehat{\mu}_{f'}(t_{\la,m})^2\widehat{\sigma}^{2'}_{f'}(t_{\la,m})}{2\widehat{\sigma}^2_{f'}(t_{\la,m})^2}\right]\frac{u}{n^\beta}\\
&\quad +\frac{1}{2} \left[\frac{\widehat{\sigma}^{2'}_{f'}(t_{\la,m})^2}{2\widehat{\sigma}^2_{f'}(t_{\la,m})^2}-\frac{\widehat{\sigma}^{2''}_{f'}(t_{\la,m})}{2\widehat{\sigma}^2_{f'}(t_{\la,m})}+\frac{2\widehat{\mu}_{f'}(t_{\la,m})\widehat{\mu}_{f'}'(t_{\la,m})\widehat{\sigma}^{2'}_{f'}(t_{\la,m})}{\widehat{\sigma}^2_{f'}(t_{\la,m})^2}\right.\\
& \qquad\qquad -\frac{\widehat{\mu}_{f'}'(t_{\la,m})^2+\widehat{\mu}_{f'}(t_{\la,m})\widehat{\mu}_{f'}''(t_{\la,m})}{\widehat{\sigma}^2_{f'}(t_{\la,m})}\\
&\qquad\qquad \left.-\frac{1}{2}\widehat{\mu}_{f'}(t_{\la,m})^2\left(\frac{2\widehat{\sigma}_{f'}^{2'}(t_{\la,m})^2}{\widehat{\sigma}^2_{f'}(t_{\la,m})^3}-\frac{\widehat{\sigma}_{f'}^{2''}(t_{\la,m})}{\widehat{\sigma}^2_{f'}(t_{\la,m})^2}\right)\right]\frac{u^2}{n^{2\beta}}\\
&\quad +\frac{1}{6}R_3(\xi)\frac{u^3}{n^{3\beta}}.
\end{align}
Based on the rates given by \eqref{eq:mu.hat}, \eqref{eq:mu.hat.prime}, \eqref{eq:sigma.t.hat} and \eqref{eq:higher.sigma.t.hat}, we obtain
\begin{align}
|\widehat{\mu}_{f'}(t_{\la,m})| \lesssim n^{-\beta}(\log n)^{-a}& ,\quad |\widehat{\mu}^{(k)}_{f'}(t_{\la,m})| \lesssim 1,\\
|\widehat{\sigma}^{2}_{f'}(t_{\la,m})| \lesssim n^{-1}\varphi_{11,m} &,\quad |\widehat{\sigma}^{2(k)}_{f'}(t_{\la,m})| \lesssim {n^{-\frac{1}{2}-\beta}}(\log n)^{-\frac 12-a},
\end{align}
for $1\leq k\leq 3$. Further calculation gives $R_3(\xi)\lesssim \frac{\widehat{\sigma}^{2''}_{f'}(\xi)}{\widehat{\sigma}^{2}_{f'}(\xi)^2}=O\left(n^{\frac{3}{2}-\beta}(\log n)^{-\frac 12-a}\varphi_{11,m}^{-2}\right)$. Substituting these into the above $\Lambda(t_{\la,m},\frac{u}{n^\beta})$ yields 
\begin{align}
\Lambda(t_{\la,m},\frac{u}{n^\beta}) &=-\frac{\widehat{\mu}_{f'}(t_{\la,m})\widehat{\mu}_{f'}'(t_{\la,m})}{n^{\beta}\widehat{\sigma}^2_{f'}(t_{\la,m})}u-\frac{\widehat{\mu}_{f'}'(t_{\la,m})^2}{2n^{2\beta}\widehat{\sigma}^2_{f'}(t_{\la,m})}u^2+o(n^{\frac{3}{2}-4\beta}\varphi_{11,m}^{-2})\\
&=-\frac{n^{1-\beta}\widehat{\mu}_{f'}(t_{\la,m})\widehat{\mu}_{f'}'(t_{\la,m})}{n\widehat{\sigma}^2_{f'}(t_{\la,m})}u-\frac{n^{1-2\beta}\widehat{\mu}_{f'}'(t_{\la,m})^2}{2n\widehat{\sigma}^2_{f'}(t_{\la,m})}u^2+o(n^{\frac{3}{2}-4\beta}\varphi_{11,m}^{-2})\\
&=n^{1-2\beta}\varphi_{11,m}^{-1}\left\{-\frac{\widehat{\mu}_{f'}'(t_{\la,m})^2u^2}{2n\varphi_{11,m}^{-1}\widehat{\sigma}^2_{f'}(t_{\la,m})}-\frac{n^\beta\widehat{\mu}_{f'}(t_{\la,m})^2u}{n\varphi_{11,m}^{-1}\widehat{\sigma}^2_{f'}(t_{\la,m})}\right\}+o(1),
\end{align}
when $n^{\frac{3}{2}-4\beta}\varphi^{-2}_{11,m}=O(1)$.

Then we study the convergence of $\mu_{n,m}$ and $\sigma^2_{n,m}$. According to \eqref{eq:mu.hat} and \eqref{eq:mu.hat.prime}, we have
\begin{equation}
|n^\beta\widehat{\mu}_{f'}(t_{\la,m})| \lesssim (\log n)^{-a},
\end{equation}
\begin{equation}
|\widehat{\mu}_{f'}'(t_{\la,m})-f_\la''(t_{\la,m})| \lesssim n^{-\beta}(\log n)^{-a}.
\end{equation}
In view of Lemma~\ref{lem:post.mode}, Assumption A2, and Assumption C, we obtain that $\widehat{\mu}_{f'}'(t_{\la,m})$ converges to $f_0''(t_m)$, and thus is bounded away from zero and infinity for sufficiently large $n$. Therefore,
\begin{align}
|\mu_{n,m}|=\bigg|\frac{n^\beta\widehat{\mu}_{f'}(t_{\la,m})}{\widehat{\mu}_{f'}'(t_{\la,m})}\bigg|\asymp |n^\beta\widehat{\mu}_{f'}(t_{\la,m})| \lesssim (\log n)^{-a}.
\end{align}

From \eqref{eq:sigma.t.hat} we have
\begin{equation}
\bigg|n\varphi_{11,m}^{-1}\widehat{\sigma}^2_{f'}(t_{\la,m})-\sigma^2\bigg| \lesssim {n^{\frac{1}{2}-2\beta}(\log n)^{-1-a}}\varphi_{11,m}^{-1}.
\end{equation}
Therefore,
\begin{align}
&\ \bigg|\sigma_{n,m}^2-\frac{\sigma^2}{f_\la''(t_{\la,m})^2}\bigg|
= \bigg|\frac{n\varphi_{11,m}^{-1}\widehat{\sigma}^2_{f'}(t_{\la,m})}{\widehat{\mu}_{f'}'(t_{\la,m})^2}- \frac{\sigma^2}{f_\la''(t_{\la,m})^2}\bigg|\\
\asymp &\ \bigg|f_\la''(t_{\la,m})^2 n\varphi_{11,m}^{-1}\widehat{\sigma}^2_{f'}(t_{\la,m})- \widehat{\mu}_{f'}'(t_{\la,m})^2\sigma^2\bigg|\\
\lesssim&\  \bigg|f_\la''(t_{\la,m})^2 \left[n\varphi_{11,m}^{-1}\widehat{\sigma}^2_{f'}(t_{\la,m})-\sigma^2\right]\bigg|+\bigg| \left[\widehat{\mu}_{f'}'(t_{\la,m})^2-f_\la''(t_{\la,m})^2\right]\sigma^2\bigg|\\
\lesssim &\ {n^{\frac{1}{2}-2\beta}(\log n)^{-1-a}}\varphi_{11,m}^{-1}+n^{-\beta}(\log n)^{-a}\\
\label{eq:sigma.conv.rate1}\lesssim&\ n^{\frac{1}{2}-2\beta}(\log n)^{-\frac 32}\varphi_{11,m}^{-1}.
\end{align}

On the other hand,
\begin{equation} \label{eq:sigma.conv.rate2}
\bigg|\frac{\sigma^2}{f_\la''(t_{\la,m})^2}-\frac{\sigma^2}{f_0''(t_m)^2}\bigg|\asymp |f_0''(t_m)^2-f_\la''(t_{\la,m})^2|\lesssim \left[f_0''(t_m)-f_\la''(t_{\la,m})\right].
\end{equation}
Since $K\in C^8(\mX,\mX)$, we have $f_\la\in C^4(\mX)$. Then the mean value theorem gives
\begin{equation}
f_0''(t_m)-f_\la''(t_{\la,m})=f_0''(t_m)-f_\la''(t_m)-f'''_\la(\xi)(t_{\la,m}-t_m).
\end{equation}
By Assumption C and Lemma~\ref{lem:post.mode} we have
\begin{equation}\label{eq:sigma.conv.rate3}
|f_0''(t_m)-f_\la''(t_{\la,m})|\lesssim \la^{r_2}+\la^{r_1} \leq 2\la^r.
\end{equation}
Combining \eqref{eq:sigma.conv.rate1}, \eqref{eq:sigma.conv.rate2} and \eqref{eq:sigma.conv.rate3}, we obtain
\begin{equation}
\bigg|\sigma_{n,m}^2-\frac{\sigma^2}{f_0''(t_m)^2}\bigg|\lesssim \la^r.
\end{equation}
This completes the proof.

\subsection{Proof of Theorem~\ref{thm:bvm}}
We first present a technical lemma and leave its proof to Section~\ref{sec:technical.lemma}. 

\begin{lem}\label{lem:tech.lemma1}
Suppose Assumption B1 holds and let $\la=n^{-\frac{1}{2}+\beta}(\log n)^{\frac 12 + a}$ for some $\frac 14<\beta<\frac12$ and $a>0$. Under event $A_n$, there exists $C>0$ such that
\begin{equation}
\Bigg|n^{-\frac{1}{2}}\varphi_{11,m}^{\frac 12}{\ell(t_{\la,m})}\bigg/\exp\left(-n^{1-2\beta}\varphi_{11,m}^{-1}\frac{\mu_{n,m}^2}{2\sigma_{n,m}^2}\right)-\frac{C}{|f_0''(t_m)|\sigma_m^{*}}\Bigg| \lesssim n^{\frac{1}{2}-2\beta}(\log n)^{-1-a}.
\end{equation}
\end{lem}

For any $x\geq 0$, define the error function as
\begin{equation}
\erf(x)=\frac{2}{\sqrt{\pi}}\int_0^x e^{-t^2}dt.
\end{equation}
By changing of variable, we have
\begin{equation}\label{eq:erf}
\int_{-A}^{B}a\exp\left(-a^2\frac{(u+b)^2}{c}\right)du=\sqrt{\frac{\pi c}{2}}\left[\erf\left(\frac{a(A-b)}{\sqrt{c}}\right) + \erf\left(\frac{a(B+b)}{\sqrt{c}}\right)\right],
\end{equation}
where $a,c,A,B>0$ and $b\in\RR$.

\subsubsection{Proof of (i)}
The proof will follow three steps.

\textbf{Step 1:} According to Theorem 1.3 in \cite{devroye2018total} and Lemma~\ref{lem:post.mode}, we have that for any $z\in\RR$,
\begin{align}
&\ \bigg|\sum_{m=1}^{M}\pi_m\Phi(z\mid t_{\la,m},n^{-1}\varphi_{11,m}\sigma_m^{*2}) - \sum_{m=1}^{M}\pi_m\Phi(z\mid t_m,n^{-1}\varphi_{11,m}\sigma_m^{*2})\bigg|\\
\leq&\ d_{TV}\left(\sum_{m=1}^{M}\pi_m \phi(\cdot \mid t_{\la,m},n^{-1}\varphi_{11,m}\sigma_m^{*2}),\ \sum_{m=1}^{M}\pi_m \phi(\cdot \mid t_m,n^{-1}\varphi_{11,m}\sigma_m^{*2})\right)\\
\lesssim &\ \sum_{m=1}^{M}\pi_m|t_{\la,m}-t_m|\lesssim \lambda^{r_1}=o(1),
\end{align}
where $d_{TV}$ is the total variation distance between two distributions. 
Thus, we only need to show
\begin{equation}
\bigg|\Pi_n(t\leq z\mid X,\bm y) - \sum_{m=1}^{M}\pi_m\Phi(z\mid t_{\la,m},n^{-1}\varphi_{11,m}\sigma_m^{*2})\bigg|\rightarrow 0
\end{equation}
for any $z\in\RR$ in $\PP_0$-probability.

\textbf{Step 2:} We work under the high probability event $A_n$ henceforth in this proof, that is, all convergence rates and bounding integrals only hold under $A_n$. 

Define a sequence of functions
\begin{equation}\label{def.m.n}
\tilde{h}_n(t)=\sum_{m=1}^{M}\ell(t_{\la,m})\tilde{\phi}_{n,m}(t)\pi(t_m),
\end{equation}
where 
\begin{equation}
\tilde{\phi}_{n,m}(t)=\exp\left(-\frac{\left(t-t_{\la,m}\right)^2}{2n^{-1}\varphi_{11,m}\sigma_m^{*2}}+n^{1-2\beta}\varphi_{11,m}^{-1}\frac{\mu_{n,m}^2}{2\sigma_{n,m}^2}\right).
\end{equation}
In this step, we will prove that
\begin{equation}\label{eq:result.1}
\bigg| \int_{-\infty}^z\ell(t)\pi(t)dt- \int_{-\infty}^z \tilde h_n(t)dt\bigg|\rightarrow 0
\end{equation}
for any $z\in\RR$. That is, $\tilde{h}_n(t)$ approximates the unnormalized limit density where each mixture component is properly rescaled. In line with the LAN condition \eqref{eq:LAN}, we expand $\tilde{h}_n(t)$ at $t=t_{\la,m}+{u}/{n^\beta}$ for $m=1,\ldots,M$, transforming $\tilde{\phi}_{n,m}(t)$ to
\begin{equation}
\nu_{n,m}(u)=\exp\left(n^{1-2\beta}\varphi_{11,m}^{-1}\left(-\frac{u^2}{2\sigma_m^{*2}}+\frac{\mu_{n,m}^2}{2\sigma_{n,m}^2}\right)\right).
\end{equation}

We consider three cases for $z$: (1) $z\leq 0$, (2) $0< z \leq 1$, and (3) $z > 1$. 

\textbf{Case (1)} $(z\leq 0)$. Since $\ell(t)=0$ in $\RR\backslash[0,1]$, the left hand side of \eqref{eq:result.1} becomes 
\begin{align}
\int_{-\infty}^{z}\tilde h_n(t)dt&\leq\sum_{m=1}^{M}\int_{-\infty}^{z}\ell(t_{\la,m})\tilde{\phi}_{n,m}(t)\pi(t_m)dt\\
&=\sum_{m=1}^{M}\int_{L_{n,m}}n^{-\beta}\ell(t_{\la,m})\nu_{n,m}(u)\pi(t_m)du
\end{align}
where we let $t=t_{\lambda,m}+u/n^\beta$ and $L_{n,m}=(-\infty, (z-t_{\la,m})n^\beta]$. By Lemma~\ref{lem:tech.lemma1} and \eqref{eq:erf} , we have
\begin{align}\label{eq:step.2.result}
\int_{L_{n,m}}n^{-\beta}\ell(t_{\la,m})\nu_{n,m}(u)\pi(t_m)du &\lesssim \int_{L_{n,m}} n^{\frac 12-\beta}\varphi_{11,m}^{-\frac 12} \exp\left(-n^{1-2\beta}\varphi_{11,m}^{-1}\frac{u^2}{2\sigma_m^{*2}}\right)du\\
&= \sqrt{\frac{\pi\sigma_m^{*2}}{2}} \left[1- \erf\left(\frac{\sqrt{n}(t_{\lambda,m}-z)}{\sqrt{2\sigma_m^{*2}\varphi_{11,m}}}\right)\right] \\
&\lesssim \exp\left(-\frac{n(t_{\lambda,m}-z)^2}{2\sigma_m^{*2}\varphi_{11,m}}\right),
\end{align}
where we use the well known inequality that $1-\erf(x)\leq \frac{2}{\sqrt{\pi}}e^{-\frac{x^2}{2}}$. Therefore, there holds that for $z \leq 0,$
\begin{equation} \label{eq:case.i.1}
\int_{-\infty}^{z}\tilde h_n(t)dt\lesssim M\exp\left(-\frac{n(t_{\lambda,1}-z)^2}{4\sigma_m^{*2}\varphi_{11,m}}\right)\rightarrow 0.
\end{equation}

\textbf{Case (2)} $(0< z \leq 1)$. Now the left hand side of \eqref{eq:result.1} becomes
\begin{equation}
\bigg| \int_{-\infty}^{z}\ell(t)\pi(t)dt- \int_{-\infty}^{z} \tilde h_n(t)dt\bigg|\leq \int_{-\infty}^0 \tilde h_n(t)dt + \bigg| \int_{0}^{z}\ell(t)\pi(t)dt- \int_{0}^{z} \tilde h_n(t)dt\bigg|.
\end{equation}
Taking $z = 0$ in~\eqref{eq:case.i.1} gives that $\int_{-\infty}^0 \tilde h_n(t)dt \rightarrow 0.$ We next bound the second term. We divide $[0,1]$ into $M$ disjoint intervals (up to overlapping endpoints that do not affect estimates of integrals), each of which centering around $t_{\la,m}$: 
\begin{equation}
[0,1]=\bigcup_{m=1}^{M}I_{n,m}, \quad I_{n,m}=[t_{\la,m}-\xi_{n,m-1},t_{\la,m}+\xi_{n,m}],\quad m=1,\ldots,M,
\end{equation}
where $\xi_{n,0}=t_{\la,1}$, $\xi_{n,m}=(t_{\la,m+1}-t_{\la,m})/2$ for $m=1,\ldots, M-1$, and $\xi_{n,M}=1-t_{\la,M}$. 
Suppose $z\in I_{n,m_0}$ for some $1\leq m_0\leq M$ and let $I'_{n,m_0}=[t_{\la,m_0}-\xi_{n,m_0-1},z]$. By the triangle inequality, we have
\begin{align}
\bigg|\int_0^{z}\ell(t)\pi(t)dt-\int_0^{z}\tilde h_n(t)dt\bigg|=& \bigg| \left(\sum_{m=1}^{m_0-1}\int_{I_{n,m}}+\int_{I'_{n,m_0}}\right) \left[\ell(t)\pi(t)-\tilde h_n(t)\right]dt\bigg|\\
\label{eq:step.one.1}\leq& \sum_{m=1}^{m_0-1}\bigg|\int_{I_{n,m}} \left[\ell(t)\pi(t)-\ell(t_{\la,m})\tilde{\phi}_{n,m}(t)\pi(t_m)\right]dt\bigg|\\
\label{eq:step.one.2}\quad&+ \sum_{m=1}^{m_0-1}\int_{[0,z]\backslash I_{n,m}}\ell(t_{\la,m})\tilde{\phi}_{n,m}(t)\pi(t_m)dt\\
\label{eq:step.one.3}\quad&+ \bigg|\int_{I'_{n,m_0}}\left[\ell(t)\pi(t)-\ell(t_{\la,m})\tilde{\phi}_{n,m}(t)\pi(t_m)\right]dt\bigg|\\
\label{eq:step.one.4}\quad&+\int_{[0,z]\backslash I'_{n,m_0}}\ell(t_{\la,m})\tilde{\phi}_{n,m}(t)\pi(t_m)dt.
\end{align}
Again, after changing of variable with $t=t_{\la,m}+{u}/{n^\beta}$, each term in \eqref{eq:step.one.1} becomes
\begin{align}
&\ \bigg|\int_{I_{n,m}}\left[\ell(t)\pi(t)-\ell(t_{\la,m})\tilde{\phi}_{n,m}(t)\pi(t_m)\right]dt\bigg|\\
=&\ \bigg|\int_{J_{n,m}}\left[\ell(t_{\la,m}+u/n^\beta)\pi(t_{\la,m}+u/n^\beta)-\ell(t_{\la,m})\nu_{n,m}(u)\pi(t_m)\right]n^{-\beta}du\bigg|\\
=&\ \bigg|\int_{J_{n,m}}n^{-\beta}\ell(t_{\la,m})\left[Z_{n,m}(u)\pi(t_{\la,m}+u/n^\beta)-\nu_{n,m}(u)\pi(t_m)\right]du\bigg|,
\end{align}
where $Z_{n,m}(u)={\ell(t_{\la,m}+u/n^\beta)}/{\ell(t_{\la,m})}$ and $J_{n,m}=[-n^\beta\xi_{n,m-1},n^\beta\xi_{n,m}]$. Applying the triangle inequality yields an upper bound of the preceding display: 
\begin{align}
&\ \bigg|\int_{J_{n,m}}n^{-\beta}\ell(t_{\la,m})Z_{n,m}(u)[\pi(t_{\la,m}+u/n^\beta)-\pi(t_m)]du\bigg|\\
&+\bigg|\int_{J_{n,m}}n^{-\beta}\ell(t_{\la,m})[Z_{n,m}(u)-\nu_{n,m}(u)]\pi(t_m)du\bigg|\\
=&\  I_1+I_2.
\end{align}
By Lemma~\ref{lem:tech.lemma1} and Theorem~\ref{thm:LAN}, we have
\begin{align}
I_1& \lesssim\bigg|\int_{J_{n,m}}n^{\frac 12-\beta}\varphi_{11,m}^{-\frac 12} \exp\left(-n^{1-2\beta}\varphi_{11,m}^{-1}\frac{\mu_{n,m}^2}{2\sigma_{n,m}^2}\right)\exp\left(n^{1-2\beta}\varphi_{11,m}^{-1}\left(-\frac{(u+\mu_{n,m})^2}{2\sigma_{n,m}^{2}}+\frac{\mu_{n,m}^2}{2\sigma_{n,m}^2}\right)\right)\\
&\qquad\qquad \cdot[\pi(t_{\la,m}+u/n^\beta)-\pi(t_m)]du\bigg|\\
&\lesssim \bigg|\int_{J_{n,m}}n^{\frac 12-\beta}\varphi_{11,m}^{-\frac 12}\exp\left(-n^{1-2\beta}\varphi_{11,m}^{-1}\frac{(u+\mu_{n,m})^2}{2\sigma_{n,m}^2}\right)[\pi(t_{\la,m}+u/n^\beta)-\pi(t_m)]du\bigg|\\
&\lesssim |t_{\la,m}-t_m|\int_{J_{n,m}}n^{\frac 12-\beta}\varphi_{11,m}^{-\frac 12}\exp\left(-n^{1-2\beta}\varphi_{11,m}^{-1}\frac{(u+\mu_{n,m})^2}{2\sigma_{n,m}^2}\right)du\\
&\quad +\bigg|\int_{J_{n,m}}n^{\frac 12-2\beta}\varphi_{11,m}^{-\frac 12}u\exp\left(-n^{1-2\beta}\varphi_{11,m}^{-1}\frac{(u+\mu_{n,m})^2}{2\sigma_{n,m}^2}\right)du\bigg|\\
&=I_{11}+I_{12}.
\end{align}
In view of \eqref{eq:mu.ni}, \eqref{eq:sigma.ni}, \eqref{eq:erf} and Lemma~\ref{lem:post.mode}, it follows that
\begin{equation}
I_{11}\lesssim |t_{\la,m}-t_m|\cdot \sqrt{2\pi \sigma_{n,m}^2}\lesssim |t_{\la,m}-t_m|\lesssim \lambda^{r_1},
\end{equation}
and
\begin{align}
I_{12}&= n^{-\beta}\bigg|\int_{J_{n,m}'}n^{\frac 12-\beta}\varphi_{11,m}^{-\frac 12}u\exp\left(-n^{1-2\beta}\varphi_{11,m}^{-1}\frac{u^2}{2\sigma_{n,m}^2}\right)du\\
&\qquad \qquad -\mu_{n,m}\int_{J_{n,m}'}n^{\frac 12-\beta}\varphi_{11,m}^{-\frac 12}\exp\left(-n^{1-2\beta}\varphi_{11,m}^{-1}\frac{u^2}{2\sigma_{n,m}^2}\right)du\bigg|\\
&\lesssim n^{-\beta}\left[2\int_0^{n^\beta(\xi_{n,m-1}\vee\xi_{n,m})}n^{\frac 12-\beta}\varphi_{11,m}^{-\frac 12}u\exp\left(-n^{1-2\beta}\varphi_{11,m}^{-1}\frac{u^2}{2\sigma_{n,m}^2}\right)du\right.\\
&\qquad\qquad  \left.+|\mu_{n,m}|\int_{J_{n,m}}n^{\frac 12-\beta}\varphi_{11,m}^{-\frac 12}\exp\left(-n^{1-2\beta}\varphi_{11,m}^{-1}\frac{(u+\mu_{n,m})^2}{2\sigma_{n,m}^2}\right)du\right]\\
&\lesssim n^{-\beta}\left[ \sigma_{n,m}^2\varphi_{11,m}^{\frac 12} \left(1-\exp\left(-{\frac{n(\xi_{n,m-1}\vee\xi_{n,m})^2}{2\sigma_{n,m}^2\varphi_{11,m}}}\right)\right)n^{-\frac 12+\beta}+\mu_{n,m}\sqrt{2\pi \sigma_{n,m}^2}\right]\\ 
&\lesssim \sqrt{\frac{\varphi_{11,m}}{n}} \wedge n^{-\beta},
\end{align}
where $J_{n,m}'=[-n^\beta\xi_{n,m-1}+\mu_{n,m},n^\beta\xi_{n,m}+\mu_{n,m}]$. Hence, $I_1\lesssim \la^{r_1} \wedge \sqrt{\frac{\varphi_{11,m}}{n}} \wedge n^{-\beta}\rightarrow 0$ under Assumption B2. By Lemma~\ref{lem:tech.lemma1} and Theorem~\ref{thm:LAN}, we have
\begin{align}
I_2 &\lesssim \bigg|\int_{J_{n,m}} n^{\frac{1}{2}-\beta}\varphi_{11,m}^{-\frac 12} \exp\left(-n^{1-2\beta}\varphi_{11,m}^{-1}\frac{\mu_{n,m}^2}{2\sigma_{n,m}^2}\right)\left[\exp\left(n^{1-2\beta}\varphi_{11,m}^{-1}\left(-\frac{(u+\mu_{n,m})^2}{2\sigma_{n,m}^{2}}+\frac{\mu_{n,m}^{2}}{2\sigma_{n,m}^{2}}\right)\right)\right.\\
&\qquad\qquad  \left.-\exp\left(n^{1-2\beta}\varphi_{11,m}^{-1}\left(-\frac{u^2}{2\sigma_m^{*2}}+\frac{\mu_{n,m}^2}{2\sigma_{n,m}^2}\right)\right)\pi(t_m)\right]du\bigg|\\
&= \bigg|\int_{J_{n,m}} n^{\frac{1}{2}-\beta}\varphi_{11,m}^{-\frac 12}\left[ \exp\left(-n^{1-2\beta}\varphi_{11,m}^{-1}\frac{(u+\mu_{n,m})^2}{2\sigma_{n,m}^{2}}\right)-\exp\left(-n^{1-2\beta}\varphi_{11,m}^{-1}\frac{u^2}{2\sigma_m^{*2}}\right)\right]du\bigg|\\
&\leq \bigg|\int_{J_{n,m}} n^{\frac{1}{2}-\beta}\varphi_{11,m}^{-\frac 12}\left[\exp\left(-n^{1-2\beta}\varphi_{11,m}^{-1}\frac{(u+\mu_{n,m})^2}{2\sigma_{n,m}^{2}}\right)-\exp\left(-n^{1-2\beta}\varphi_{11,m}^{-1}\frac{u^2}{2\sigma_{n,m}^2}\right)\right]du\bigg|\\
&\quad+ \bigg|\int_{J_{n,m}} n^{\frac{1}{2}-\beta}\varphi_{11,m}^{-\frac 12}\left[\exp\left(-n^{1-2\beta}\varphi_{11,m}^{-1}\frac{u^2}{2\sigma_{n,m}^{2}}\right)-\exp\left(-n^{1-2\beta}\varphi_{11,m}^{-1}\frac{u^2}{2\sigma_m^{*2}}\right)\right]du\bigg|\\
&=I_{21}+I_{22}.
\end{align}
Without loss of generality we assume $\mu_{n,m}\geq 0$. Then, combining \eqref{eq:mu.ni}, \eqref{eq:sigma.ni} and\eqref{eq:erf} gives that
\begin{align}
I_{21}&=\Bigg| \sqrt{\frac{\pi\sigma_{n,m}^2}{2}}\left[\erf\left(\frac{\xi_{n,m-1}n^{\frac 12}-\mu_{n,m}n^{\frac 12-\beta}}{\sqrt{2\sigma_{n,m}^2\varphi_{11,m}}}\right) + \erf\left(\frac{\xi_{n,m}n^{\frac 12}+\mu_{n,m}n^{\frac 12-\beta}}{\sqrt{2\sigma_{n,m}^2\varphi_{11,m}}}\right)\right] \\
&\qquad -\sqrt{\frac{\pi \sigma_{n,m}^2}{2}}\left[\erf\left(\frac{\xi_{n,m-1}n^{\frac 12}}{\sqrt{2\sigma_{n,m}^2\varphi_{11,m}}}\right)+\erf\left(\frac{\xi_{n,m}n^{\frac 12}}{\sqrt{2\sigma_{n,m}^2\varphi_{11,m}}}\right)\right]\Bigg|\\
&\lesssim\Bigg| \erf\left(\frac{\xi_{n,m-1}n^{\frac 12}-\mu_{n,m}n^{\frac 12-\beta}}{\sqrt{2\sigma_{n,m}^2\varphi_{11,m}}}\right)-\erf\left(\frac{\xi_{n,m-1}n^{\frac 12}}{\sqrt{2\sigma_{n,m}^2\varphi_{11,m}}}\right)\Bigg| \\
&\qquad + \Bigg| \erf\left(\frac{\xi_{n,m}n^{\frac 12}+\mu_{n,m}n^{\frac 12-\beta}}{\sqrt{2\sigma_{n,m}^2\varphi_{11,m}}}\right)-\erf\left(\frac{\xi_{n,m}n^{\frac 12}}{\sqrt{2\sigma_{n,m}^2\varphi_{11,m}}}\right)\Bigg|\\
&\leq \mu_{n,m}n^{\frac 12-\beta} \cdot \exp\left(- \frac{\xi_{n,m-1}^2n}{2\sigma_{n,m}^2\varphi_{11,m}}\right) + \mu_{n,m}n^{\frac 12-\beta} \cdot \exp\left(- \frac{\xi_{n,m}^2n}{2\sigma_{n,m}^2\varphi_{11,m}}\right)\\
&\lesssim e^{-cn\varphi_{11,m}^{-1}}
\end{align}
for some $c>0$. In view of \eqref{eq:sigma.ni}, \eqref{eq:erf} and Theorem~\ref{thm:LAN}, we obtain
\begin{align}
I_{22} &=\bigg|\sqrt{\frac{\pi \sigma_{n,m}^2}{2}}\left[\erf\left(\frac{\xi_{n,m-1}n^{\frac 12}}{\sqrt{2\sigma_{n,m}^2\varphi_{11,m}}}\right)+\erf\left(\frac{\xi_{n,m}n^{\frac 12}}{\sqrt{2\sigma_{n,m}^2\varphi_{11,m}}}\right)\right]\\
&\qquad -\sqrt{\frac{\pi \sigma_m^{*2}}{2}}\left[\erf\left(\frac{\xi_{n,m-1}n^{\frac 12}}{\sqrt{2\sigma_m^{*2}\varphi_{11,m}}}\right)+\erf\left(\frac{\xi_{n,m}n^{\frac 12}}{\sqrt{2\sigma_m^{*2}\varphi_{11,m}}}\right)\right]\bigg|\lesssim |\sigma_m^{*2}-\sigma_{n,m}^2|\lesssim \lambda^r.
\end{align}
Therefore, $I_2 \rightarrow 0$.

Similarly, by changing of variable and Lemma~\ref{lem:tech.lemma1}, each term in \eqref{eq:step.one.2} becomes
\begin{align}
I_3&=\int_{[0,z]\backslash I_{n,m}}\ell(t_{\la,m})\tilde{\phi}_{n,m}(t)\pi(t_m)dt\\
&=\int_{K_{n,m}}n^{-\beta}\ell(t_{\la,m})\nu_{n,m}(u)\pi(t_m)du\\
&\lesssim \int_{K_{n,m}}n^{\frac{1}{2}-\beta}\varphi_{11,m}^{-\frac 12}\exp\left(-n^{1-2\beta}\varphi_{11,m}^{-1}\frac{\mu_{n,m}^2}{2\sigma_{n,m}^2}\right)\exp\left(n^{1-2\beta}\varphi_{11,m}^{-1}\left(-\frac{u^2}{2\sigma_m^{*2}}+\frac{\mu_{n,m}^2}{2\sigma_{n,m}^2}\right)\right)du\\
&= \int_{K_{n,m}}n^{\frac{1}{2}-\beta}\varphi_{11,m}^{-\frac 12}\exp\left(-n^{1-2\beta}\varphi_{11,m}^{-1}\frac{u^2}{2\sigma_m^{*2}}\right)du,
\end{align}
where $K_{n,m}=[-n^\beta t_{\la,m},-n^\beta\xi_{n,m-1}]\cup[n^\beta\xi_{n,m}, n^\beta(z-t_{\la,m})]$. It then follows that
\begin{equation}
I_3\lesssim n^\beta \cdot n^{\frac{1}{2}-\beta}\varphi_{11,m}^{-\frac 12}\exp\left(-n^{1-2\beta}\varphi_{11,m}^{-1}\frac{n^{2\beta}(z-t_{\la,m})^2}{2\sigma_m^{*2}}\right)\rightarrow 0.
\end{equation}
Following the same arguments, we can show that \eqref{eq:step.one.3} and \eqref{eq:step.one.4} converge to zero. This proves~\eqref{eq:result.1} for $0 < z \leq 1$.

\textbf{Case (3)} ($z > 1$). From Case (2) we can see that
\begin{equation}
\bigg|\int_{-\infty}^1\ell(t)\pi(t)dt-\int_{-\infty}^1\tilde h_n(t)dt\bigg|\rightarrow 0.
\end{equation}
Note that $\ell(t)=0$ in $\RR\backslash[0,1]$. Using similar arguments as in Case (1), it holds that $\int_1^{z} \tilde h_n(t)dt \rightarrow 0,$ proving~\eqref{eq:result.1} for $z > 1$.

\textbf{Step 3:} We normalize $\tilde h_n(t)$ to a density 
\begin{align}
h_n(t)=\frac{\tilde h_n(t)}{\int_\RR \tilde h_n(t)dt}.
\end{align}
Note that \eqref{eq:result.1} implies 
\begin{equation}\label{eq:result.2}
\bigg|\left(\int_\RR \ell(t)\pi(t)dt\right)^{-1}- \left(\int_\RR \tilde h_n(t)dt\right)^{-1}\bigg|\rightarrow 0.
\end{equation}
Hence, for any $z \in \RR$, we have 
\begin{align}
\bigg| \int_{-\infty}^{z} \pi_n(t\mid X,\bm y)dt - \int_{-\infty}^{z}h_n(t)du\bigg| &\leq \left(\int_\RR \ell(t)\pi(t)dt\right)^{-1}\bigg| \int_{-\infty}^{z}\ell(t)\pi(t)dt-\int_{-\infty}^{z}\tilde h_n(t) dt \bigg|\\
&\quad +\int_{-\infty}^{z} \bigg|\left(\int_\RR  \ell(t)\pi(t)dt\right)^{-1}-\left(\int_\RR  \tilde h_n(t)dt\right)^{-1}\bigg|\tilde h_n(t)dt\\
\label{step.3.result1}&\rightarrow 0,
\end{align}
where the last line follows from \eqref{eq:result.1} and \eqref{eq:result.2}. 

Rewrite $\tilde h_n(t)$ defined in \eqref{def.m.n} to
\begin{equation}
\tilde h_n(t)=\sum_{m=1}^{M}\pi(t_m)\ell(t_{\la,m})\exp\left(n^{1-2\beta}\varphi_{11,m}^{-1}\frac{\mu_{n,m}^2}{2\sigma_{n,m}^2}\right) {\sqrt{2\pi n^{-1}\varphi_{11,m}\sigma_m^{*2}}}\phi(t\mid t_{\la,m},n^{-1}\varphi_{11,m}\sigma_m^{*2}),
\end{equation}
which is a linear combination of $\phi(t\mid t_{\la,m},n^{-1}\varphi_{11,m}\sigma_m^{*2}).$ Hence, the density function after normalization is 
\begin{equation}
h_n(t)=\sum_{m=1}^{M}\tilde{\pi}_{n,m}\phi(t\mid t_{\la,m},n^{-1}\varphi_{11,m}\sigma_m^{*2}),
\end{equation}
with weights
\begin{align}
\tilde{\pi}_{n,m}&=\frac{\pi(t_m)\ell(t_{\la,m})\exp\left(n^{1-2\beta}\varphi_{11,m}^{-1}\frac{\mu_{n,m}^2}{2\sigma_{n,m}^2}\right){\sqrt{2\pi n^{-1}\varphi_{11,m}\sigma_m^{*2}}}}{\sum_{m=1}^{M}\pi(t_m)\ell(t_{\la,m})\exp\left(n^{1-2\beta}\varphi_{11,m}^{-1}\frac{\mu_{n,m}^2}{2\sigma_{n,m}^2}\right){\sqrt{2\pi n^{-1}\varphi_{11,m}\sigma_m^{*2}}}}\\
&=\frac{\pi(t_m)\sigma_m^*n^{-\frac{1}{2}}\varphi_{11,m}^{\frac 12}\ell(t_{\la,m})/\exp\left(-n^{1-2\beta}\varphi_{11,m}^{-1}\frac{\mu_{n,m}^2}{2\sigma_{n,m}^2}\right)}{\sum_{m=1}^{M}\pi(t_m)\sigma_m^*n^{-\frac{1}{2}}\varphi_{11,m}^{\frac 12}\ell(t_{\la,m})/\exp\left(-n^{1-2\beta}\varphi_{11,m}^{-1}\frac{\mu_{n,m}^2}{2\sigma_{n,m}^2}\right)}\\
&=\frac{\pi(t_m)C|f_0''(t_m)|^{-1}+c_{n,m}}{\sum_{m=1}^{M}\pi(t_m)C|f_0''(t_m)|^{-1}+c_{n,m}}
\end{align}
where the existence of sequences $c_{n,m}=O(n^{\frac{1}{2}-2\beta}(\log n)^{-1-a})$ is guaranteed by Lemma~\ref{lem:tech.lemma1}. Hence, we arrive at 
\begin{equation}
\tilde{\pi}_{n,m} =\frac{\pi(t_m)|f_0''(t_m)|^{-1}}{\sum_{m=1}^{M}\pi(t_m)|f_0''(t_m)|^{-1}}+c_{n,m}' =: \pi_m + c_{n,m}',
\end{equation}
for some $c_{n,m}'=O(n^{\frac{1}{2}-2\beta}(\log n)^{-1-a})$. It then holds that
\begin{align}
& \bigg|\int_{-\infty}^{z}h_n(t)dt-\int_{-\infty}^{z}\sum_{m=1}^{M}\pi_m \phi(t\mid t_{\la,m},n^{-1}\varphi_{11,m}\sigma_m^{*2})dt\bigg| \\  \leq & \; \sum_{m=1}^{M}\int_{-\infty}^{z} |\tilde{\pi}_{n,m}-\pi_m|\phi(t\mid t_{\la,m},n^{-1}\varphi_{11,m}\sigma_m^{*2})dt
\rightarrow 0.\label{step.3.result2}
\end{align}
Combining \eqref{step.3.result1} and \eqref{step.3.result2}, we obtain that for any $z$,
\begin{equation}
\EE_{\PP_0}\bigg|\int_{-\infty}^{z}\pi_n(t\mid X,\bm y)dt-\int_{-\infty}^{z}\sum_{m=1}^{M}\pi_m \phi(t\mid t_{\la,m},n^{-1}\varphi_{11,m}\sigma_m^{*2})dt\bigg|\ind_{A_n}\rightarrow 0.
\end{equation}
This together with  $\mathbb{E}_{\PP_0}(\ind_{A_n^c})=\PP_0(A_n^c)\leq n^{-10}$ gives that
\begin{equation}
\bigg|\Pi_n(t\leq z\mid X,\bm y) - \sum_{m=1}^{M}\pi_m\Phi(z\mid t_{\la,m},n^{-1}\varphi_{11,m}\sigma_m^{*2})\bigg|\rightarrow 0
\end{equation}
for any $z\in\RR$ in $\PP_0$-probability. This completes the proof.

\subsubsection{Proof of (ii)}
Denote $\zeta_0=t_1$, $\zeta_m=\frac{1}{2}(t_{m+1}-t_{m})$, $m=1,\ldots, M-1$, $\zeta_M=1-t_M$. Then,
\begin{equation}
[0,1]=\bigcup_{m=1}^{M}I_m, \quad I_m=[t_m-\zeta_{m-1},t_m+\zeta_m],\quad m=1,\ldots,M.
\end{equation}
We first bound the unnormalized difference
\begin{equation}\label{eq:part.ii}
\bigg|\int_{-\infty}^z \ell(t)\ind_{I_m}(t)\pi(t)dt-\int_{-\infty}^z\ell(t_{\la,m})\tilde{\phi}_{n,m}(t)\pi(t_m)dt\bigg|
\end{equation}
under $A_n$ by considering three cases for $z$: (1) $z\leq t_m-\zeta_{m-1}$, (2) $t_m-\zeta_{m-1}< z<t_m+\zeta_m$, and (3) $z\geq t_m+\zeta_m$.

\textbf{Case 1} $(z\leq t_m-\zeta_{m-1})$. Since $z\notin I_m$, \eqref{eq:part.ii} becomes 
\begin{align}
\int_{-\infty}^z\ell(t_{\la,m})\tilde{\phi}_{n,m}(t)\pi(t_m)dt&=\int_{-\infty}^{(z-t_{\la,m})n^\beta}\ell(t_{\la,m})\nu_{n,m}(u)\pi(t_m)du\\ &\lesssim \int_{-\infty}^{(z-t_{\la,m})n^\beta} n^{\frac 12-\beta}\varphi_{11,m}^{-\frac 12} \exp\left(-n^{1-2\beta}\varphi_{11,m}^{-1}\frac{u^2}{2\sigma_m^{*2}}\right)du\\
&= \sqrt{\frac{\pi\sigma_m^{*2}}{2}} \left[1- \erf\left(\frac{\sqrt{n}(t_{\lambda,m}-z)}{\sqrt{2\sigma_m^{*2}\varphi_{11,m}}}\right)\right] \\
&\lesssim \exp\left(-\frac{n(t_{\lambda,m}-t_m-\zeta_{m-1})^2}{2\sigma_m^{*2}\varphi_{11,m}}\right).
\end{align}

\textbf{Case 2} $(t_m-\zeta_{m-1}< z< t_m+\zeta_m)$. In this case, we consider
\begin{align}
&\ \bigg|\int_{t_m-\zeta_{m-1}}^z\left[\ell(t)\pi(t)-\ell(t_{\la,m})\tilde{\phi}_{n,m}(t)\pi(t_m)\right]dt\bigg|\\
=&\ \bigg|\int_{H_{n,m}}\left[\ell(t_{\la,m}+u/n^\beta)\pi(t_{\la,m}+u/n^\beta)-\ell(t_{\la,m})\nu_{n,m}(u)\pi(t_m)\right]n^{-\beta}du\bigg|\\
=&\ \bigg|\int_{H_{n,m}}n^{-\beta}\ell(t_{\la,m})\left[Z_{n,m}(u)\pi(t_{\la,m}+u/n^\beta)-\nu_{n,m}(u)\pi(t_m)\right]du\bigg|\\
\leq&\ \bigg|\int_{H_{n,m}}n^{-\beta}\ell(t_{\la,m})Z_{n,m}(u)[\pi(t_{\la,m}+u/n^\beta)-\pi(t_m)]du\bigg|\\
&+\bigg|\int_{H_{n,m}}n^{-\beta}\ell(t_{\la,m})[Z_{n,m}(u)-\nu_{n,m}(u)]\pi(t_m)du\bigg|\\
&= I_1'+I_2',
\end{align}
where $H_{n,m}=[(t_m-t_{\la,m}-\zeta_{m-1})n^\beta,(z-t_{\la,m})n^\beta]$. Following similar arguments as used in the proof of Part (i), it can be shown that $I_1'\lesssim \la^{r_1}$ and
\begin{align}
I_2'\lesssim \mu_{n,m}n^{\frac 12-\beta} \cdot \exp\left(- \frac{(z-t_{\la,m})^2n}{2\sigma_{n,m}^2\varphi_{11,m}}\right).
\end{align}

\textbf{Case 3} $(z \geq t_m+\zeta_m)$. Again, $z\notin I_m$ and \eqref{eq:part.ii} becomes
\begin{align}
\int_{t_m+\zeta_m}^z\ell(t_{\la,m})\tilde{\phi}_{n,m}(t)\pi(t_m)dt&=\int_{(t_m-t_{\la,m}+\zeta_m)n^\beta}^{(z-t_{\la,m})n^\beta}\ell(t_{\la,m})\nu_{n,m}(u)\pi(t_m)du\\ &\lesssim \int_{(t_m-t_{\la,m}+\zeta_m)n^\beta}^\infty n^{\frac 12-\beta}\varphi_{11,m}^{-\frac 12} \exp\left(-n^{1-2\beta}\varphi_{11,m}^{-1}\frac{u^2}{2\sigma_m^{*2}}\right)du\\
&= \sqrt{\frac{\pi\sigma_m^{*2}}{2}} \left[1- \erf\left(\frac{\sqrt{n}(t_m-t_{\la,m}+\zeta_m)}{\sqrt{2\sigma_m^{*2}\varphi_{11,m}}}\right)\right] \\
&\lesssim \exp\left(-\frac{n(t_m-t_{\la,m}+\zeta_m)^2}{2\sigma_m^{*2}\varphi_{11,m}}\right).
\end{align}
Combining the three cases, we obtain that under $A_n$,
\begin{equation}
\bigg|\int_{-\infty}^z \ell(t)\ind_{I_m}(t)\pi(t)dt-\int_{-\infty}^z\ell(t_{\la,m})\tilde{\phi}_{n,m}(t)\pi(t_m)dt\bigg|\lesssim \la^{r_1}\vee n^{\frac 12-\beta} \cdot \exp\left(- \frac{(z-t_{\la,m})^2n}{2\sigma_{n,m}^2\varphi_{11,m}}\right).
\end{equation}
Let $\Pi_{n,m}(\cdot \mid X,\by)$ be the posterior of $t \ind_{I_m}$. Following the same arguments as in part (i) again, we can show that
\begin{equation}
\bigg|\Pi_{n,m}(t'\leq z\mid X,\bm y) - \Phi(z\mid t_{\la,m},n^{-1}\varphi_{11,m}\sigma_m^{*2})\bigg|\lesssim \la^{r_1}\vee n^{\frac 12-\beta} \cdot \exp\left(- \frac{(z-t_{\la,m})^2n}{2\sigma_{n,m}^2\varphi_{11,m}}\right)
\end{equation}
in $\PP_0$-probability.

By Lemma~\ref{lem:post.mode}, we have $|\hat{t}_m-b_n-t_{\la,m}|\lesssim n^{-\beta}\sqrt{\log n} \vee n^{-\beta}\log n = n^{-\beta}\log n$. Thus,
\begin{equation}
\bigg|\Pi_{n,m}(t'\leq z\mid X,\bm y) - \Phi(z\mid \hat t_m-b_n,n^{-1}\varphi_{11,m}\sigma_m^{*2})\bigg|\lesssim \la^{r_1}\vee n^{\frac 12-\beta} \cdot \exp\left(- \frac{(z-t_{\la,m})^2n}{2\sigma_{n,m}^2\varphi_{11,m}}\right) \vee n^{-\beta}\log n.
\end{equation}
Now we consider the posterior of $\sqrt{\frac{n}{\varphi_{11,m}}}(t \ind_{I_m}(t) -\hat t_m+b_n)$. By changing of variable, it follows that
\begin{align}
&\quad \; \bigg|\Pi_{n,m}'(t'\leq z\mid X,\bm y) - \Phi(z\mid 0,\sigma_m^{*2})\bigg| \\
&= \bigg|\Pi_{n,m}\left(t'\leq \sqrt{\frac{\varphi_{11,m}}{n}}z+\hat t_m-b_n\mid X,\bm y\right)-\Phi\left(\sqrt{\frac{\varphi_{11,m}}{n}}z+\hat t_m-b_n|\hat t_m-b_n,n^{-1}\varphi_{11,m}\sigma_m^{*2}\right)\bigg|\\
&\lesssim n^{\frac 12-\beta} \cdot \exp\left\{- \frac{(\sqrt{\frac{\varphi_{11,m}}{n}}z+\hat t_m-t_{\la,m}-b_n)^2n}{2\sigma_{n,1}^2\varphi_{11,m}}\right\} \\
&\lesssim n^{\frac 12-\beta} \cdot \exp \left\{-\left(  \sqrt{\frac{\varphi_{11,m}}{n}}z \vee (\hat t_m-t_{\la,m}) \vee b_n \right)^2\frac{n}{2\sigma_{n,m}^2\varphi_{11,m}}\right\}\\
&\lesssim n^{\frac 12-\beta} \cdot \exp \left\{-\frac{b_n^2n}{2\sigma_{n,m}^2\varphi_{11,m}}\right\}\\
&\lesssim n^{\frac 12-\beta} \cdot \exp \left\{-\frac{n^{1-2\beta}(\log n)^2} {2\sigma_{n,m}^2\varphi_{11,m}}\right\}\rightarrow 0.
\end{align}
This completes the proof.

\subsection{Proof of Theorem~\ref{thm:frequentist}}

\subsubsection{Proof of (i)}

Let $F_n(z) = \Pi_n(t \leq z \mid X,\bm{y})$ and $G_n(z) = \sum_{m=1}^M \pi_m \Phi(z\mid t_m, n^{-1}\varphi_{11,m}\sigma_m^{*2})$. Note that $G_n(\cdot)$ is a deterministic function, and its derivative $G_n'(\cdot)$ is the density function of a Gaussian mixture. The variance of each component distribution in $G_n$ goes to zero in view of Assumption B2 and conditions in Theorem~\ref{thm:bvm}. For sufficiently large $n$, using the analytical expression of $G_n'(\cdot)$ and elementary calculus, we can show that $G_n'(\cdot)$ has at least $M$ local modes, denoted by $t_{m,G}$, such that $t_{m,G}\rightarrow t_m$. On the other hand, $G_n'(\cdot)$ cannot have more than $M$ local modes in view of Corollary 2.4 in \cite{carreira2003number}; hence, $\{t_{m, 1}, \ldots, t_{M, G}\}$ are the only local modes of $G_n'(\cdot)$. For large enough $n$ and each $m$, we consider an interval $(t_{m,G}-\delta_m, t_{m,G} + \delta_m)$ for some $\delta_m>0$ such that $G_n''(z)>0$ when $z\in (t_{m,G}-\delta_m, t_{m,G})$ and $G_n''(z)<0$ when $z\in (t_{m,G}, t_{m,G}+\delta_m)$. 

By Theorem~\ref{thm:bvm} (i), we have $|F_n(z) - G_n(z)|\to 0$ for any $z\in \mathbb{R}$ in $\mathbb{P}_0$-probability. The following arguments and conclusions in Step 1--4 hold with $\mathbb{P}_0$-probability tending to 1 because of this convergence in $\PP_0$-probability. 

\textbf{Step 1:} We first show that there exists a $t_{m,F}$ in the neighborhood of $t_{m,G}$ such that $F_n''(t_{m,F})=0$, for $m=1,\ldots, M$. Suppose $F_n''(z) \neq 0$ for any $z\in (t_{m,G}-\delta_m, t_{m,G}+\delta_m)$, Without loss of generality we assume $F''_n(z) > 0$ when $z\in (t_{m,G}-\delta_m, t_{m,G}+\delta_m)$. Since $G_n(z)$ is concave on $(t_{m,G}, t_{m,G}+\delta_m)$, 
\begin{equation}\label{eq:concave.g}
G_n(t_{m,G}+\delta_m/2) > (G_n(t_{m,G})+G_n(t_{m,G}+\delta_m))/2+\epsilon,
\end{equation}
for some $\epsilon>0$. Since $F_n(z)$ is convex on $(t_{m,G}, t_{m,G}+\delta_m)$, 
\begin{equation}
F_n(t_{m,G}+\delta_m/2) < (F_n(t_{m,G})+F_n(t_{m,G}+\delta_m))/2.
\end{equation}
For sufficiently large $n$, it holds that with $\mathbb{P}_0$-probability tending to 1 $|F_n(z) - G_n(z)| < \epsilon/2$ for $z = t_{m,G}, t_{m,G}+\delta_m/2, t_{m,G}+\delta_m$. Therefore,
\begin{equation}
G_n(t_{m,G}+\delta_m/2) > (F_n(t_{m,G})+F_n(t_{m,G}+\delta_m))/2 + \epsilon/2> F_n(t_{m,G}+\delta_m/2) + \epsilon/2,
\end{equation}
which is a contradiction. This proves that there exists $t_{m,F} \in (t_{m,G}-\delta_m, t_{m,G}+\delta_m)$ such that $F_n''(t_{m,F}) = 0$. 

\textbf{Step 2:} We show that $t_{m,F}\rightarrow t_m$ in $\mathbb{P}_0$-probability. Suppose there exists $\delta>0$ such that $|t_{m,G} - t_{m,F}| > \delta$ for any sufficiently large $n$. Without loss of generality we assume $t_{m,G} < t_{m,F}$ and $F_n''(z) < 0$ when $z\in (t_{m,F}, t_{m,G} + \delta_m)$ and $F_n''(z) > 0$ when $z\in (t_{m,G} - \delta_m, t_{m,F})$. Thus, $G_n$ is concave on $(t_{m,G}, t_{m,F})$ while $F_n$ is convex on $(t_{m,G}, t_{m,F})$. This is a contradiction using the same argument in Step 1. Combining this with $t_{m,G}\rightarrow t_m$ shows the convergence of $t_{m, F}$. 

\textbf{Step 3:} In this step, we show that $t_{m,F}$ must be a local mode of $F_n'(z)$. Suppose that $F_n''(z) > 0$ when $z\in (t_{m,F}, t_{m,G} + \delta_m)$ and $F_n''(z) < 0$ when $z\in (t_{m,G} - \delta_m, t_{m,F})$, yielding
\begin{equation}
F_n(t_{m,F}+\delta_m/2) < (F_n(t_{m,F})+F_n(t_{m,F}+\delta_m))/2.
\end{equation}
For sufficiently large $n$, it holds with $\mathbb{P}_0$-probability tending to 1 that $|F_n(z) - G_n(z)| < \epsilon/4$ for $x = t_{m,G}, t_{m,G}+\delta_m/2, t_{m,G}+\delta_m$. Invoking \eqref{eq:concave.g},
\begin{equation}
G_n(t_{m,G}+\delta_m/2) > (F_n(t_{m,G})+F_n(t_{m,G}+\delta_m))/2 + 3\epsilon/4.
\end{equation}
For sufficiently large $n$, it holds with $\mathbb{P}_0$-probability tending to 1 that $|F_n(z_1) - F_n(z_2)| < \epsilon/4$ for $z_1 = t_{m,G}$, $z_2 = t_{m_F}$, $z_1 = t_{m,G}+\delta_m/2$, $z_2 = t_{m,F}+\delta_m/2$ and $z_1 = t_{m,G}+\delta_m$, $z_2 = t_{m,F}+\delta_m$. Therefore,
\begin{equation}
G_n(t_{m,G}+\delta_m/2) > (F_n(t_{m,F})+F_n(t_{m,F}+\delta_m))/2 + \epsilon/2 > F_n(t_{m,F}+\delta_m/2) + \epsilon/2.
\end{equation}
However,
\begin{equation}
G_n(t_{m,G}+\delta_m/2) < F_n(t_{m,G}+\delta_m/2) + \epsilon/4 < F_n(t_{m,F}+\delta_m/2) + \epsilon/2,
\end{equation}
which is a contradiction. This completes Step 3. 

\textbf{Step 4:} In the last step, we show that the number of local modes of $F_n'(z)$ is exactly $M$. We have proven that $F_n'(\cdot)$ has at least $M$ local modes. Suppose that there exists $t_{m',F}\in (0,1)$ and $\delta_{m'}>0$ such that $t_{m',F}$ is a local mode of $F_n'(z)$ and $G_n''(z)\neq 0$ for $z\in (t_{m',F} - \delta_{m'}, t_{m',F} + \delta_{m'})$ for any sufficiently large $n$. Without loss of generality assume $G_n''(z)>0$ for $z\in (t_{m',F} - \delta_{m'}, t_{m',F} + \delta_{m'})$. Thus, on $(t_{m',F} - \delta_{m'}, t_{m',F} + \delta_{m'})$, $G_n(z)$ is convex while $F_n(z)$ is concave. By similar arguments used in Step 1, we can obtain a contradiction. Hence, the number of local modes of $F_n'(\cdot)$ is exactly $M$. 

This completes the proof.

\subsubsection{Proof of (ii)}

By Taylor expansion of $\widehat \mu_{f'}$, we obtain  
\begin{equation}
\widehat \mu_{f'}(t) = \widehat \mu_{f'}(t_m) + (t - t_m) \widehat \mu_{f'}'(\xi)
\end{equation}
for some $\xi$ between $t$ and $t_m$. Since $\hat t_m$ is a local extremum of $\hat \mu_f$, there holds $\widehat \mu_{f'}(\hat t_m) = 0$. Substituting $t = \hat t_m$ into the expansion above yields 
\begin{equation}
\widehat \mu_{f'}(t_m) + (\hat t_m - t_m) \widehat \mu_{f'}'(\xi) = 0.
\end{equation}
Lemma~\ref{thm:nonasy.bound} and Assumption C ensure that $\widehat \mu_{f'}'(x) \overset{p}{\to} f_0''(x)$, and Lemma~\ref{lem:post.mode} implies that $\hat t_m \overset{p}{\to} t_m$. Therefore, $\widehat \mu_{f'}'(\xi) \overset{p}{\to} f_0''(t_m)$, and thus $\widehat \mu_{f'}'(\xi)$ is bounded away from zero and infinity in view of Assumption A3. It thus follows that
\begin{equation}\label{taylor.t.m}
\hat t_m - t_m = -\frac{\widehat \mu_{f'}(t_m) }{\widehat \mu_{f'}'(\xi) }.
\end{equation}
Let $\Delta_{n}(\cdot) = K_{10}(\cdot, X) [K(X, X) + n \lambda \bm I_n]^{-1} f_0(X)$. Conditioning on $X$, it holds that
\begin{align}
\widehat{\mu}_{f'}(t_m) &= K_{10}(t_m, X) [K(X, X) + n \lambda \bm I_n]^{-1} \bm y \\
& \sim N \left( \Delta_{n}(t_m), \sigma^2 K_{10}(t_m, X) [K(X, X) + n \lambda \bm I_n]^{-2} K_{10}(t_m, X)^T \right).
\end{align}
Hence,
\begin{equation}
\frac{\widehat{\mu}_{f'}(t_m) - \Delta_{n}(t_m)}{\sigma \sqrt{K_{10}(t_m, X) [K(X, X) + n \lambda \bm I_n]^{-2} K_{10}(t_m, X)^T}} \bigg| X \sim N(0,1),
\end{equation}
which implies that 
\begin{equation}
\frac{\widehat{\mu}_{f'}(t_m) - \Delta_{n}(t_m)}{\sigma \sqrt{K_{10}(t_m, X) [K(X, X) + n \lambda \bm I_n]^{-2} K_{10}(t_m, X)^T}}  \sim N(0,1).
\end{equation}
By Slutsky's theorem, we obtain
\begin{align}
\sqrt{\frac{1}{K_{10}(t_m, X) [K(X, X) + n \lambda \bm I_n]^{-2} K_{10}(t_m, X)^T}}\left[\hat t_{m} - t_m  + \frac{\Delta_{n}(t_m)}{\widehat{\mu}'_{f'}(t_m)}\right] &\overset{d}{\to} \\
& N\left(0, \sigma^2 f_0''(t_m)^{-2} \right).
\end{align}
Note that
\begin{align}
&\ \frac{K_{10}(\hat t_m, X) [K(X, X) + n \lambda \bm I_n]^{-2} K_{10}(\hat t_m, X)^T}{K_{10}(t_m, X) [K(X, X) + n \lambda \bm I_n]^{-2} K_{10}(t_m, X)^T} - 1\\
= & \ \frac{K_{10}(\hat t_m, X)[K(X, X) + n \lambda \bm I_n]^{-2} (K_{10}(\hat t_m, X)^T - K_{10}(t_m, X))}{K_{10}(t_m, X) [K(X, X) + n \lambda \bm I_n]^{-2} K_{10}(t_m, X)^T }\\
&+ \frac{(K_{10}(\hat t_m, X)^T - K_{10}(t_m, X))[K(X, X) + n \lambda \bm I_n]^{-2}  K_{10}(t_m, X))}{K_{10}(t_m, X) [K(X, X) + n \lambda \bm I_n]^{-2} K_{10}(t_m, X)^T}.
\end{align}
Consider the eigendecomposition of $K(X, X) = Q_n\Lambda_n Q_n^T$, where $\Lambda_n = \text{diag}(u_1,\ldots,u_n)$ and $Q_n^T = Q_n^{-1}$. Denote $(p_1,\ldots,p_n) = K_{10}(t_m, X) Q_n$, likewise $(q_1,\ldots,q_n) = K_{10}(\hat t_m, X) Q_n$. Then
\begin{align}
K_{10}(t_m, X) [K(X, X) + n \lambda \bm I_n]^{-2} K_{10}(t_m, X)^T &= K_{10}(t_m, X) Q_n \Lambda_n^{-2} Q_n^T K_{10}(t_m, X)\\
&=\s \frac{p_i^2}{(u_i+n\lambda)^2}.
\end{align}
By the Cauchy–Schwarz inequality, we have
\begin{align}
& K_{10}(\hat t_m, X)[K(X, X) + n \lambda \bm I_n]^{-2} (K_{10}(\hat t_m, X)^T - K_{10}(t_m, X)) \\ = & \s \frac{q_i (q_i-p_i)}{(u_i+n\lambda)^2}
\leq \sqrt{\s \frac{q_i^2}{(u_i+n\lambda)^2} \s \frac{(q_i-p_i)^2}{(u_i+n\lambda)^2}}.
\end{align}
Since $K_{10}(\hat t_m, X_i) - K_{10}(t_m, X_i) \overset{p}{\to} 0$ uniformly for $1\leq i \leq n$, we have
\begin{equation}
\s \frac{q_i^2}{(u_i+n\lambda)^2} \bigg/ \s \frac{p_i^2}{(u_i+n\lambda)^2} \overset{p}{\to} 1
\end{equation}
and
\begin{equation}
\s \frac{(q_i-p_i)^2}{(u_i+n\lambda)^2} \bigg/ \s \frac{p_i^2}{(u_i+n\lambda)^2} \overset{p}{\to} 0.
\end{equation}
Hence, it follows that 
\begin{equation}
\frac{K_{10}(\hat t_m, X)[K(X, X) + n \lambda \bm I_n]^{-2} (K_{10}(\hat t_m, X)^T - K_{10}(t_m, X))}{K_{10}(t_m, X) [K(X, X) + n \lambda \bm I_n]^{-2} K_{10}(t_m, X)^T } \overset{p}{\to} 0.
\end{equation}
Similarly, it can be shown that
\begin{equation}
\frac{(K_{10}(\hat t_m, X)^T - K_{10}(t_m, X))[K(X, X) + n \lambda \bm I_n]^{-2}  K_{10}(t_m, X))}{K_{10}(t_m, X) [K(X, X) + n \lambda \bm I_n]^{-2} K_{10}(t_m, X)^T} \overset{p}{\to} 0.
\end{equation}
Therefore,
\begin{equation}
\frac{K_{10}(\hat t_m, X) [K(X, X) + n \lambda \bm I_n]^{-2} K_{10}(\hat t_m, X)^T}{K_{10}(t_m, X) [K(X, X) + n \lambda \bm I_n]^{-2} K_{10}(t_m, X)^T} \overset{p}{\to} 1.
\end{equation}
Recall that $\widehat \mu_{f'}'(\hat t_m) \overset{p}{\to} f_0''(t_m)$. Therefore,
by Slutsky's theorem again, we arrive at 
\begin{align}
\frac{\sigma |\widehat\mu_{f'}'(\hat t_m)|}{\sqrt{K_{10}(\hat t_m, X) [K(X, X) + n \lambda \bm I_n]^{-2} K_{10}(\hat t_m, X)^T}}\left[\hat t_{m} - t_m  + \frac{\Delta_{n}(t_m)}{\widehat{\mu}'_{f'}(t_m)}\right] \overset{d}{\to} N\left(0, 1\right).
\end{align}
Hence, an asymptotic $1-\alpha$ confidence interval of $t_m + \Delta_n(t_m)/f_0''(t_m)$ is
\begin{equation}
\hat t_m \pm z_{\alpha/2} \frac{ \sigma \sqrt{K_{10}(\hat t_m, X) [K(X, X) + n \lambda \bm I_n]^{-2} K_{10}(\hat t_m, X)^T}}{|\widehat\mu_{f'}'(\hat t_m)| }.
\end{equation}
This completes the proof.

\subsection{Proof of Theorem~\ref{thm:example.poly}}
For any $j,l\leq s$, we have
\begin{equation}
|K^{\alpha,s}_{jl}(x,x')|=\bigg|\s \mu_i \psi_i^{(j)}(x)\psi_i^{(l)}(x')\bigg| \lesssim \s i^{-2\alpha +j+l},
\end{equation}
which is finite when $\alpha>\frac{j+l+1}{2}$.
Thus, Assumption B1 holds when $s\geq 4$ and $\alpha> 9/2$. According to Lemma 11 in \cite{liu2023estimation}, when $\alpha>\frac{j+l+1}{2}$, we have
\begin{equation}
\sup_{x\in \mX}|\varphi_{jl}(x)|=\sup_{x\in \mX}\bigg|\s \frac{\mu_i}{\la+\mu_i}\psi^{(j)}(x)\psi^{(l)}(x)\bigg|\lesssim \s \frac{\mu_i i^{j+l}}{\mu_i+\la} \asymp \la^{-\frac{j+l+1}{2\alpha}}.
\end{equation}
Hence, Assumption B2 is satisfied when $\alpha> 3$. In view of Lemma~1, Lemma~11 and Lemma~13 in \cite{liu2023estimation}, when $\alpha>k+1/2$, we have 
\begin{equation}
\|f_\lambda^{(k)}-f_0^{(k)}\|_\infty \lesssim\lambda^{\frac{1}{2}-\frac{k}{2\alpha}}.
\end{equation}
This verifies Assumption C with $r_1=\frac{\alpha-1}{2\alpha}$,  $r_2=\frac{\alpha-2}{2\alpha}$ when $\alpha>5/2$. Finally, by Assumption E we have $\varphi_{11}(x)=\s \frac{\mu_i}{\la+\mu_i}\psi_i'(x)^2\rightarrow \infty$ as $\la\rightarrow 0$. Thus, a sufficient condition for the boundedness of $n^{\frac{3}{2}-4\beta}\varphi^{-2}_{11,m}$ in Theorem~\ref{thm:LAN} and~\ref{thm:bvm} is $n^{\frac{3}{2}-4\beta}=O(1)$, which implies that $\beta\geq \frac 38.$
This completes the proof.

\subsection{Proof of Lemma~\ref{lem:exp.deriv.deterministic}}
Let $f_0=\s f_i\psi_i$. Then, for any $k\leq s$,
\begin{equation}
f_\la^{(k)}-f_0^{(k)}=-\s \frac{\la}{\la+\mu_i}f_i\psi_i^{(k)}.
\end{equation}
Hence,
\begin{equation}
\| f_\la^{(k)}-f_0^{(k)} \|_\infty \leq \s \frac{\la}{\la+\mu_i}|f_i|\cdot i^k\lesssim \la^{\frac 12 - \frac{k}{2e\gamma}}\s \frac{\la^{\frac 12 + \frac{k}{2e\gamma}}\cdot e^{-\gamma i}i^k}{\la+e^{-2\gamma i}}e^{\gamma i}|f_i|.
\end{equation}
Note that $i^k\leq e^{\frac{ki}{e}}$, then by Young's inequality for products, we have
\begin{equation}
\la^{\frac 12 + \frac{k}{2e\gamma}}\cdot e^{-\gamma i}i^k\leq \la^{\frac 12 + \frac{k}{2e\gamma}}\cdot e^{(\frac ke-\gamma)i}\leq \left(\frac 12 + \frac{k}{2e\gamma}\right) \la + \left(\frac 12 - \frac{k}{2e\gamma}\right) e^{-2\gamma i}\leq \la + e^{-2\gamma i}.
\end{equation}
Therefore, $\| f_\la^{(k)}-f_0^{(k)} \|_\infty \lesssim \la^{\frac 12 - \frac{k}{2e\gamma}} \s e^{\gamma i}|f_i|\lesssim \la^{\frac 12 - \frac{k}{2e\gamma}}$. This completes the proof.

\subsection{Proof of Theorem~\ref{thm:example.exp}}

It is easy to see that $K_{\gamma,s} \in C^8(\mX,\mX)$ for any $\gamma >0$ and $s\geq 4$; thus Assumption B1 is satisfied.

Note that
\begin{equation}
\sup_{x\in \mX}|\varphi_{jl}(x)|=\bigg|\s \frac{\mu_i}{\la+\mu_i}\psi_i^{(j)}(x)\psi_i^{(l)}(x)\bigg|\lesssim \s \frac{e^{-2\gamma i}i^{j+l}}{\la+e^{-2\gamma i}}\leq \frac{e^{-2\gamma i}e^{\frac{(j+l)i}{e}}}{\la+e^{-2\gamma i}}.
\end{equation}
By Young's inequality for products, when $2e\gamma >j+l$ we have
\begin{equation}
\la^{\frac{j+l}{2e\gamma}}\cdot e^{-2\gamma i+\frac{(j+l)i}{e}}\leq \left(1-\frac{j+l}{2e\gamma}\right) \la + \left(\frac{j+l}{2e\gamma}\right) e^{-2\gamma i}\leq \la + e^{-2\gamma i}.
\end{equation}
Hence, $\varphi_{jl}\lesssim \la^{-\frac{j+l}{2e\gamma}}$ for $2e\gamma >j+l$ and Assumption B2 holds for $\gamma>\frac{5}{2e}$. In view of Lemma~\ref{lem:exp.deriv.deterministic}, we have Assumption C satisfied with $r_1=r_2=\frac{e\gamma-2}{2e\gamma}$ and $\gamma >\frac{2}{e}$.

Finally, since $\la=o(1)$, we have $\varphi_{11,m}^{-2}=o(1)$ under Assumption E. Thus, a sufficient condition for the boundedness of $n^{\frac{3}{2}-4\beta}\varphi^{-2}_{11,m}$ is $\beta\geq \frac 38$. This completes the proof.

\subsection{Proof of Lemma~\ref{lem:tech.lemma1}}
\label{sec:technical.lemma}

The likelihood function \eqref{eq:likelihood} gives
\begin{align}
n^{-\frac{1}{2}}\varphi_{11,m}^{\frac12}{\ell(t_{\la,m})}&=\frac{C}{\sqrt{n\varphi_{11,m}^{-1}\widehat{\sigma}^2_{f'}(t_{\la,m})}}\exp\left(-\frac{\widehat{\mu}_{f'}(t_{\la,m})^2}{2\widehat{\sigma}^2_{f'}(t_{\la,m})}\right)\\
&=\frac{C}{\sqrt{n\varphi_{11,m}^{-1}\widehat{\sigma}^2_{f'}(t_{\la,m})}}\exp\left(-n^{1-2\beta}\varphi_{11,m}^{-1}\frac{\mu_{n,m}^2}{2\sigma_{n,m}^2}\right),
\end{align}
where $\mu_{n,m}$ and $\sigma^2_{n,m}$ are defined in Theorem~\ref{thm:LAN}. Note that
\begin{align}
\bigg|\frac{1}{\sqrt{n\varphi_{11,m}^{-1}\widehat{\sigma}^2_{f'}(t_{\la,m})}}-\frac{1}{|f_0''(t_m)|\sigma_m^*}\bigg|&\asymp ||f_0''(t_m)|\sigma_m^*-\sqrt{n\varphi_{11,m}^{-1}\widehat{\sigma}^2_{f'}(t_{\la,m})}|\\
&\asymp |f_0''(t_m)^2\sigma_m^{*2}-n\varphi_{11,m}^{-1}\widehat{\sigma}^2_{f'}(t_{\la,m})|.
\end{align}
Substituting $f_0''(t_m)^2\sigma_m^{*2}=\sigma^2$ into the right side yields 
\begin{align}
\bigg|f_0''(t_m)^2\sigma_m^{*2}-n\varphi_{11,m}^{-1}\widehat{\sigma}^2_{f'}(t_{\la,m})\bigg| = \bigg| n\varphi_{11,m}^{-1}\widehat{\sigma}^2_{f'}(t_{\la,m})-\sigma^2\bigg| \lesssim n^{\frac{1}{2}-2\beta}(\log n)^{-1-a}.
\end{align}
This completes the proof.

\section{Additional simulation results}
\subsection{Effect of noise standard derivation and credible level}
We carried out additional experiments to investigate the effect of noise standard derivation and credible levels $1 - \alpha$. 
We used the same regression function shown in the paper and generated more noisy data by increasing the noise standard deviation $\sigma$ from 0.1 to 0.2. As expected, results worsen, particularly for smaller sample sizes. This is because the GP tends to produce more wiggly curves. For example, looking at the percentages of correctly estimating $M$ for $\alpha=.05$, calculated over 100 replicated datasets, we observed the following results: for $n=100$ we obtained 19\% and 85\% for Beta (1,1) and Beta(2,3), respectively, versus 47\% and 86\% of Figure 3 in the paper; for $n=500$ we obtained 52\% and 94\% for Beta (1,1) and Beta(2,3), respectively, versus 95\% and 99\% for $\sigma=0.1$. We notice that, as already shown in the main simulation, the $Beta(2, 3)$ prior and larger sample sizes help identifying the correct number of local extrema. 

Next, we used this additional simulation study to investigate the performance of HPDR for different values of $\alpha$. Results for sample sizes $n=100$, $n=500$ and $n = 1000$ and the two Beta priors are reported in the two tables below. For each combination of prior and sample size, we generated 100 simulated datasets.  

\begin{table}[ht]
\small
\centering
\begin{tabular}{lllllll}
\hline
$Beta(1, 1)$ & $\alpha = 0.001$ & $\alpha = 0.005$ & $\alpha = 0.01$ & $\alpha = 0.03$ & $\alpha = 0.05$ & $\alpha = 0.1$\\ 
\hline
$n=100$ & 56\% & 53\% & 51\%& 18\% & 19\% & 35\%  \\ 
$n=500$ & 25\%  &  32\% & 34\% & 43\% & 52\% & 60\% \\ 
$n=1000$ & 27\% &  35\% & 44\% & 57\% & 76\% & 84\%  \\ 
\hline
\end{tabular}
\caption{$Beta(1, 1)$.  Percentages of correctly estimated number of $t$'s. The results are calculated on 100 simulated data.}
\label{tbl:beta_1_1_percent}
\end{table}

\begin{table}[ht]
\small
\centering
\begin{tabular}{lllllll}
\hline
$Beta(2, 3)$ & $\alpha = 0.001$ & $\alpha = 0.005$ & $\alpha = 0.01$ & $\alpha = 0.03$ & $\alpha = 0.05$ & $\alpha = 0.1$\\ 
\hline
$n=100$ & 87\% & 88\% & 88\% & 86\% & 85\% & 77\% \\ 
$n=500$ & 95\% & 96\% &  95\% &  95\% &  94\% & 89\%\\ 
$n=1000$ &  95\%  & 95\% &  96\% &  94\% & 94\% & 95\% \\ 
\hline
\end{tabular}
\caption{$Beta(2, 3)$.  Percentages of  correctly estimated number of $t$'s. The results are calculated on 100 simulated data.}
\label{tbl:beta_2_3_percent}
\end{table}

\begin{figure}[H]
\centering
\includegraphics[width = 0.6\textwidth]{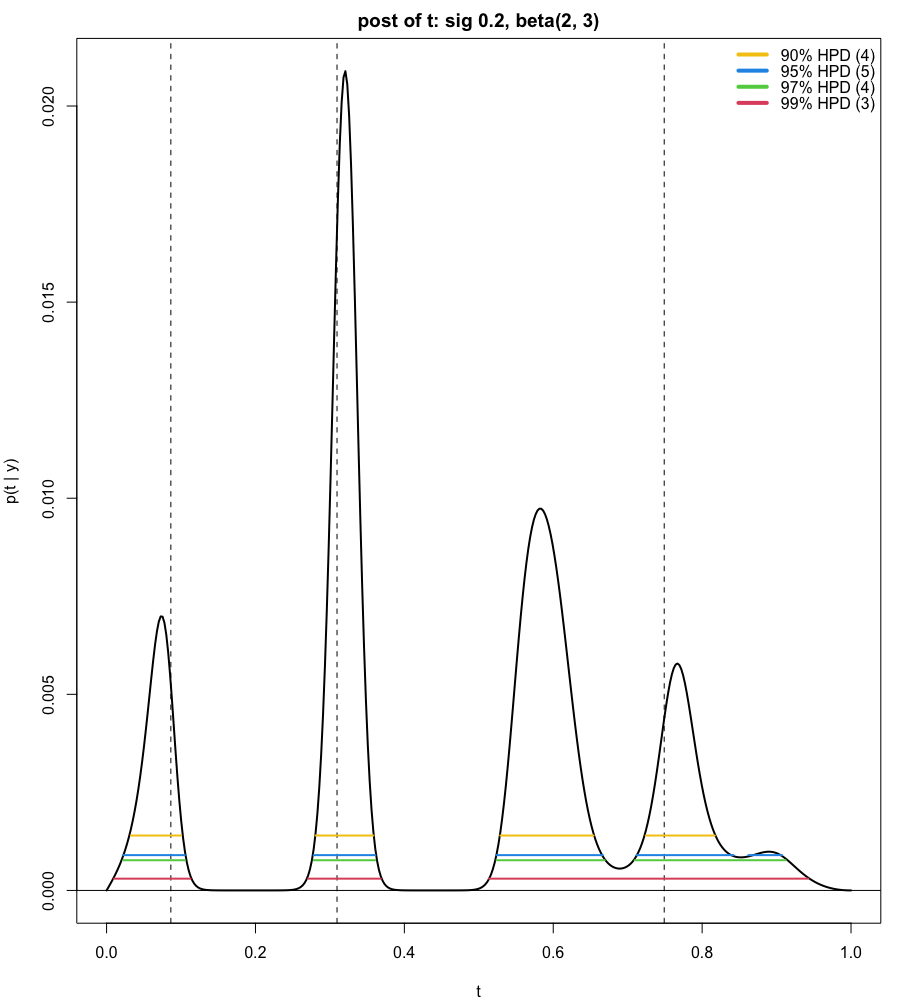}
\caption{Effect of $\alpha$ on the estimated number of local extrema. The posterior density function is based on one simulated dataset with $n = 100$.}
\label{fig:hpd_alpha}
\end{figure}

In this additional study, we observed that increasing values of $\alpha$ did not necessarily correspond to larger estimated numbers of local extrema. This is because situations like the one shown in Figure \ref{fig:hpd_alpha} can occur. Therefore, larger or smaller $\alpha$ values do not necessarily imply more or fewer separated HPDR segments. Overall, results confirm the fairly robust estimation performance of the Beta(2,3) prior in estimating $M$. 

\subsection{Highly fluctuated regression function with large $M$}

Upon suggestion from one of the reviewers, we performed a new simulation using the regression function $\sin(k \pi x)$ for $x \in [0, 1]$, and assessed how the estimated number of local extrema converges to the true $M$. 
We considered $k = 10$ and $k = 100$ with varying $n$; with this regression function, the true number of local extrema is $M = k$. Other simulation configurations mirrored the main paper's setup, including the noise standard deviation, observed $x$ values, and the number of replications. The proposed method is implemented using the same settings as in the simulation study in the main paper, unless otherwise stated.

We observe that when $k = 10$, our method is able to correctly estimate $M$ 77\% of the time even with sample size as small as 30. This percentage increases steadily to (93\%, 99\%, 100\%) as $n$ increases to $(200, 300, 500)$, respectively.

\begin{figure}[h]
\centering
\includegraphics[width=.7\linewidth]{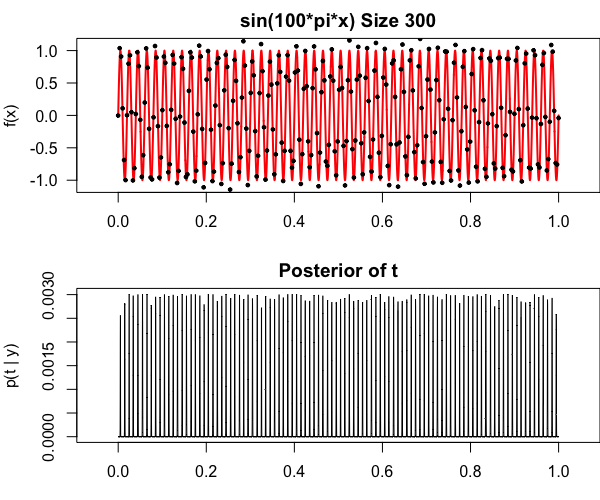}
\caption{Data (top) and the posterior density of $t$ (bottom) when $f(x) = \sin(100\pi x)$ (red curve in the top plot). Results are based on one simulated dataset with sample size $n = 300$.}
\label{fig:postM100}
\end{figure}

When $k = 100$, $M$ is correctly estimated only 4\% of the time when $n = 300$ (compared to 99\% when $k = 10$), indicating the challenge of large $k = 100$. We have looked into this challenging scenario and found that for this highly fluctuated function, even simpler tasks such as function estimation become challenging. For example, the model struggles to distinguish between a highly fluctuated function and a flat function when $n = 300$, which is not surprising as indicated in the top plot of Figure~\ref{fig:postM100}. This has prompted us to find an effective strategy for this challenging function in which we incorporate the shape of the function into guided hyperparameter tuning. If we have prior knowledge that there are many local extrema, we can confine the hyperparameter searching space, ruling out some basins of the marginal likelihood that do not result in the regression shape being interested. For example, setting the upper bound when searching for $(h, \lambda)$ to (0.1, 0.0001) as opposed to (10000, 10000)  used in our default implementation, leads to the results reported in Table~\ref{table:Mhat.k100}, which show a substantially improved estimation of $M$. For example, the proposed method can estimate the correct value of $M$ with $n = 300$ in all 100 simulations. The posterior distribution of $t$ in one simulation when $k = 100$ is shown in Figure \ref{fig:postM100}. In this simulation, which is typical across 100 replications, our method correctly identifies the number and location of 100 local extrema. We acknowledge that prior information on the shape of the unknown function might not always be available.

\begin{table}[ht]
\centering
\begin{tabular}{cccccc}
\hline
& 70-79 & 80-89 & 90-99 & 100 & $>100$ \\
\hline
$n = 200$    & 18    & 76    & 6    & 0   & 0    \\
$n = 225$     & 0     & 1     & 14    & 85  & 0    \\
$n = 250$     & 0     & 0     & 0     & 99  & 1    \\
$n = 300$    & 0     & 0     & 0     & 100 & 0    \\
\hline
\end{tabular}
\caption{Frequency of $\hat{M}$ falling in each interval when $k = 100$. Results are based on 100 repeated simulations. } \label{table:Mhat.k100}
\end{table}

\end{document}